\documentclass[aps,prb,reprint,amsmath,amssymb,superscriptaddress,twocolumn,nofootinbib]{revtex4-2}
\usepackage{graphicx}
\usepackage{xcolor}
\usepackage{subfigure}
\usepackage{hyperref,hypcap}
\usepackage{braket}
\usepackage{amsmath}
\usepackage{commath}
\usepackage{mathtools}
\usepackage{chemformula}

\begin{document}
	
	\title{Staggered spin Hall conductivity}
	
	\author{Fei Xue} 
	\affiliation{Physical Measurement Laboratory, National Institute of Standards and Technology, Gaithersburg, MD 20899, USA}
	\affiliation{Institute for Research in Electronics and Applied Physics \& Maryland Nanocenter,	University of Maryland, College Park, MD 20742}
	\author{Paul M. Haney}	
	\affiliation{Physical Measurement Laboratory, National Institute of Standards and Technology, Gaithersburg, MD 20899, USA}
	
	\date{\today}
	
	\begin{abstract}
		The intrinsic spin Hall effect plays an important role in spintronics applications, such as spin-orbit torque-based memory. The bulk space group symmetry determines the form of the bulk spin current conductivity tensor.  This paper considers materials for which the local point group symmetry of individual atoms is lower than the global (bulk) symmetry.  This enables a position-dependent spin current response, with additional tensor components allowed relative to the bulk response.  We present a general method to compute the position-dependent intrinsic spin Hall conductivity, with similar computational effort relative to computing the bulk spin Hall conductivity.  We also present the general symmetry-constrained form of the position-dependent spin current response.  We apply this method to 1T'-\ch{WTe2}, which exhibits a conventional spin Hall conductivity tensor component $\sigma^y_{xz}$ and a staggered unconventional component $\sigma^z_{xz}$. The magnitude of these two components, around 100 and 20 $(\rm \Omega\cdot cm)^{-1}$, respectively, are comparable to the spin-orbit torque exerted on adjacent ferromagnets in experiments. We then consider orthorhombic \ch{PbTe}, in which both uniform and staggered spin current conductivity are one order of magnitude larger.
	\end{abstract}
	
	\maketitle
	
\section{Introduction}  
\begin{figure}[htbp]
		\includegraphics[width=.95\columnwidth]{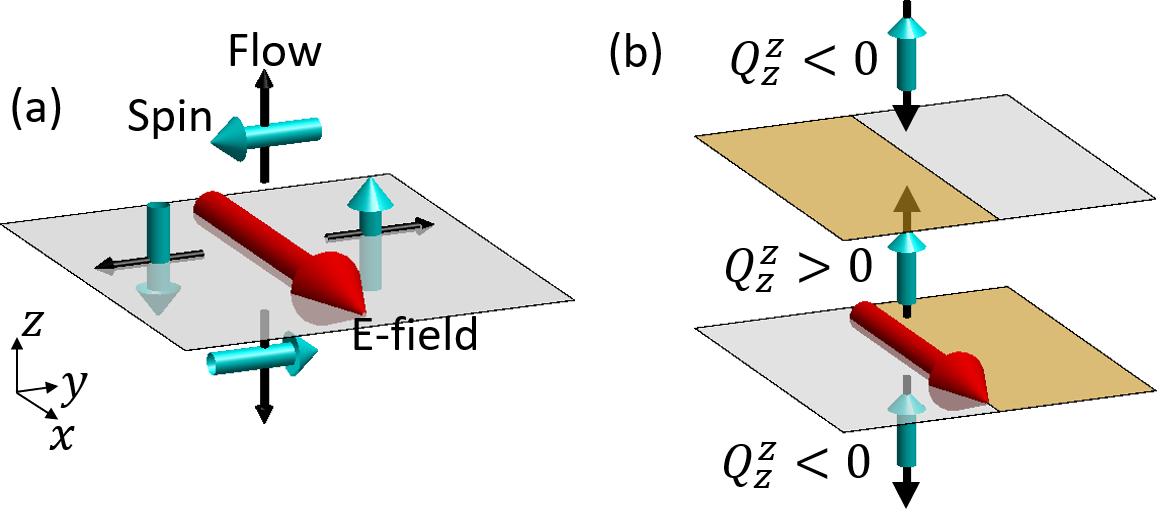}
		\caption{(a) Configuration of conventional spin Hall effect in crystals with cubic symmetry.  The applied electric field (red arrow along $x$), spin current flow direction (black arrows), and spin current spin direction (cyan arrows) are mutually orthogonal. (b) Additional unconventional spin current response for a system in which $y \rightarrow -y$ mirror symmetry is broken in an opposite sense in adjacent layers (different colors of $xy$ planes represent the broken symmetry).   }
		\label{fig:staggereshe}
	\end{figure}

In spintronics, the efficient generation of spin current is a key component to electrically controlling a system's magnetic state.  A prominent mechanism for electrically generating spin current is the spin Hall effect \cite{SinovaRMP2015}.  The spin current generated from the spin Hall effect typically flows perpendicular to the applied electric field, with spin polarization perpendicular to both flow and electric field directions (see Fig.~\ref{fig:staggereshe}(a)) \cite{Hirsch1999,Sinova2004,SinovaRMP2015}.  This prototypical spin current configuration can be understood on general symmetry grounds, and applies to materials with cubic symmetry.  The spin current generated from the spin Hall effect has been utilized in bilayer heterostructures composed of a substrate - which functions as a source of spin Hall current - and a ferromagnetic layer.  The spin current from the substrate exerts a spin transfer torque on the adjacent ferromagnet \cite{liu2011spin,miron2011perpendicular,liu2012spin,liu2012current}, an effect which belongs to the family of so-called ``spin-orbit torques''.  This mechanism for electrically controlling the ferromagnetic orientation is useful for a range of applications, including magnetic random access memory and magnetic domain wall motion \cite{manchon2019current}.  However, the symmetry-derived constraints on the form of the spin current limit its usefulness for some important applications, such as electrically switching perpendicularly magnetized layers \cite{wang2013low}.
	
	A recently developed approach to overcoming this limitation of the spin Hall effect is the use of materials or systems with reduced symmetry \cite{yu2014switching}, enabling additional components of spin current. Recent experiments \cite{MacNeill2016,MacNeill2017,stiehl2019current,Stiehl2019,shi2019all} showed that substrates with broken in-plane symmetry exert unconventional spin-orbit torques on adjacent ferromagnets.  These torques are consistent with the presence of spin current flowing in the substrate whose spin polarization is aligned to the direction of spin flow [see Fig.~\ref{fig:staggereshe}(b)].  Subsequent first principles calculations demonstrated that the computed torque in these systems is indeed mostly the consequence of this unconventional spin Hall current \cite{Xue2020SOT}.  Other measurements of nonlocal transport also indicate an unconventional spin current response in low symmetry materials \cite{Safeer2019,Song2020,zhao2020unconventional}.
	
	The general symmetry constraints on the spin Hall conductivity for different bulk symmetry groups has been established previously \cite{Seemann2015}. However, the experimental work referenced above utilizes a class of materials with a nonsymmorphic symmetry group.  In such materials, the symmetry of a specific lattice site can be lower than the global symmetry of the material.  This leads to a spatially varying response which can be staggered or ``hidden'': the system response is nonzero on individual sites, but vanishes after summing over all sites comprising a unit cell. Identifying and utilizing the hidden response of materials has been a common theme in recent work, such as computing the staggered spin-splitting present in the ground state of bulk materials \cite{zhang2014hidden}, or the staggered electrically induced spin present in CuMnAs \cite{wadley2016electrical,manchon2019current}.
	
	In this work, we present a method for computing the spatially-resolved intrinsic spin current conductivity. This method requires the same computational overhead as evaluating the bulk spin Hall conductivity.  We also present the general symmetry-constrained form of the position-dependent spin Hall conductivity. Using this approach within density functional theory, we calculate the staggered spin Hall conductivity for two materials: 1T'-\ch{WTe2} and orthorhombic \ch{PbTe}. We find that the magnitude of the unconventional staggered spin Hall conductivity in \ch{WTe2} is similar to the unconventional dampinglike torque computed \cite{Xue2020SOT} and measured\cite{MacNeill2016} in \ch{WTe2}-ferromagnet bilayers. For \ch{PbTe}, we find a staggered spin Hall conductivity response whose magnitude exceeds that of \ch{WTe2} by an order of magnitude. 
	
	The observation that the staggered spin Hall conductivity can be closely related to the unconventional spin-orbit torque exerted on adjacent ferromagnets \cite{Xue2020SOT} provides a key motivation for our work. Another motivation is that the computation of the bulk staggered spin Hall response is significantly less intensive than the direct computation of the spin-orbit torque in heterostructures.  The method we present here can be adopted easily in existing first-principles calculations, enabling a more efficient search for materials to serve as the source of dampinglike spin-orbit torque.  Finally, we note that this work focuses on intrinsic contributions to the spin Hall conductivity, with the expectation that the intrinsic mechanism dominates extrinsic mechanisms such as skew-scattering and side-jump in transition metals with strong spin-orbit coupled bands \cite{SinovaRMP2015,Nagaosa_SHE,Tanaka_SHE}.

\section{Formalism}

\subsection{Method}
We first present a method for computing the position-dependent spin current conductivity tensor within a periodic unit cell.  This method is most readily applied in the tight-binding representation, which we utilize in this work. The direct construction of the position-dependent spin current operator in a localized orbital basis is straightforward, however it generally requires additional computation overhead relative to the bulk current operator (see Appendix~\ref{App.position}). Here we show an alternative approach which does not require this additional overhead. The idea is to compute the net spin influx into each atomic site (or each atomic layer), and use this information to construct the position-dependent spin current.  
	
We consider nonmagnetic materials, where the Hamiltonian is given by:
\begin{eqnarray}
	H\left({\bf r}\right)=\frac{{\bf p}^2}{2m}+V({\bf r}) +\alpha_{\rm{SOC}}\left({\bf p}\times \nabla V({\bf r}) \right)\cdot {\bf s} \label{eq:H}~,
\end{eqnarray}
where ${\bf p}$ is the momentum operator, $V({\bf r})$ includes the crystal field potential and the electron-electron interaction energy at mean-field level, which may be spin-dependent. $\alpha_{\rm{SOC}}$ parameterizes the spin-orbit coupling and ${\bf s}$ is the spin operator.  To proceed, we consider Eq.~\ref{eq:H} where the interaction terms are evaluated in the ground state.  Eq.~\ref{eq:H} can then be expressed in a single particle tight-binding representation:
\begin{eqnarray}
	H = \sum_{\nu,\nu',\alpha,\alpha'} H_{\nu \nu'\alpha \alpha'} ~c_{\nu \alpha}^\dagger c_{\nu' \alpha'}~,
\end{eqnarray}
where $c^\dagger_{\nu\alpha}$ is the creation operator for an orbital $\alpha$ (where the orbital label also includes spin) centered on atomic site $\nu$, and $c_{\nu'\alpha'}$ is the annihilation operator for orbital $\alpha'$ on site $\nu'$. The tight-binding Hamiltonian matrix element is given by:
\begin{eqnarray}
	H_{\nu\nu'\alpha\alpha'} = \int d{\bf r} ~\phi_{\nu \alpha}^*\left({\bf r}\right) H\left({\bf r}\right)\phi_{\nu'\alpha'}\left({\bf r}\right)
\end{eqnarray}
where $\phi_{\nu\alpha}({\bf r})$ is the localized real space wave function of orbital $\alpha$ centered at site $\nu$.  The set of orbitals $\{\phi_{\nu\alpha}({\bf r})\}$ form the basis for the tight-binding representation.
	
\begin{figure}[htbp]
	\includegraphics[width=1\columnwidth]{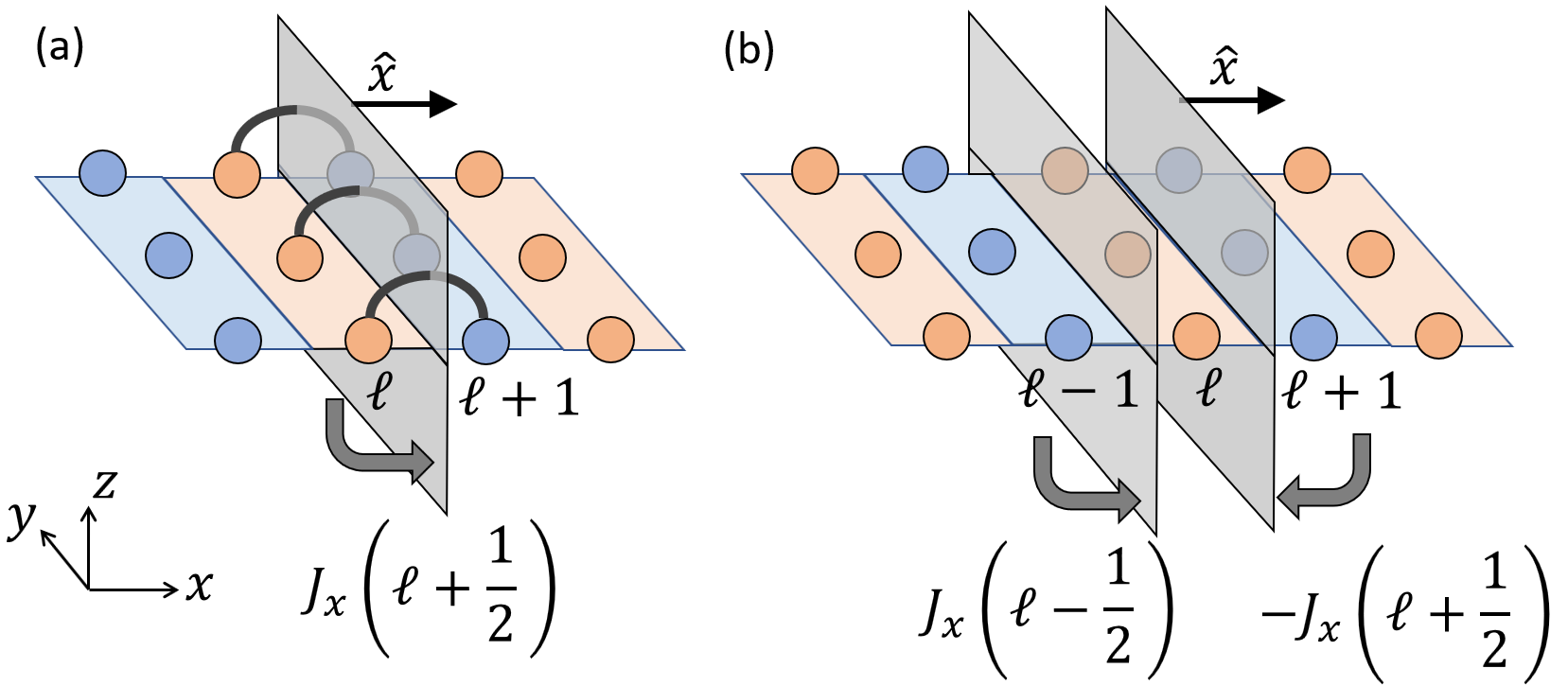}
	\caption{Schematic of a tight-binding representation of a lattice.  $\ell$ and $\ell+1$ label adjacent layers (colored blue and red, respectively). (a) $J_{x}\left(\ell+\frac{1}{2}\right)$ is the flux density passing through the plane separating layers $\ell$ and $\ell+1$.  For this example, ${\hat i} = {\hat x}$. (b) The difference between $J_{x}\left(\ell+\frac{1}{2}\right)$ and $J_{x}\left(\ell-\frac{1}{2}\right)$ is the net flux density into the layer $\ell$.}
	\label{fig:Qflux}
\end{figure}

The tight-binding model is an abstraction of the full real space description of the system.  The two primitive objects in a tight-binding description are sites, and links between sites. Site-defined quantities include number and spin.  Link-defined quantities include flux and spin flux.  In this work we focus on bulk, periodic systems.  In periodic systems, a site in the unit cell corresponds to an extended {\it layer}, which we specify with an integer label $\ell$, and the flux between two sites corresponds to a flux density $J$ passing through a {\it plane} separating layers. (Note that we use the terms flux density and current density interchangeably.) We label planes with half-integers; a plane with label $\ell+1/2$ lies between the adjacent layers $\ell$ and $\ell+1$.  We also specify the direction normal to the plane in the subscript of the flux density: $J_i$ (see Fig.~\ref{fig:Qflux}).

Our method relies on first partitioning a unit cell into layers stacked along a chosen direction $\hat{i}$.  Denoting the $i^{\rm th}$ component of a site's position as $r_i$, a layer consists of all sites with the same value of $r_i$. The number of layers $N$ is therefore equal to the number of unique values of $r_i$ among all sites in the unit cell.

\begin{figure}[htbp]
	\includegraphics[width=.95\columnwidth]{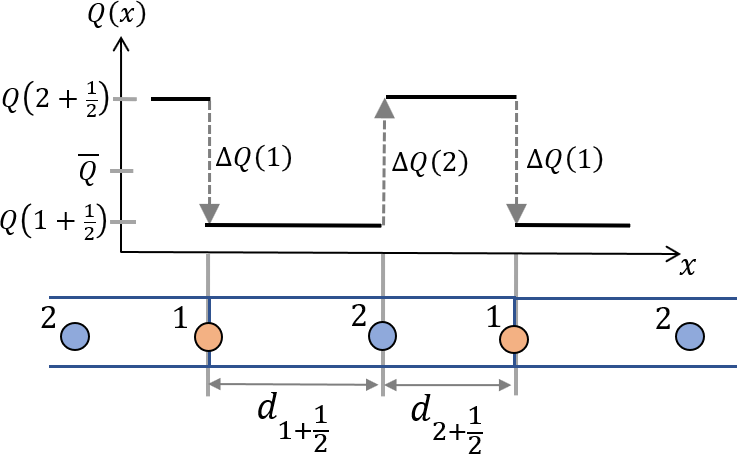}
	\caption{Schematic of spin current distribution $Q^{\alpha}(x)$ for a system with 2 sites (layers) in a unit cell.  $\Delta Q^{\alpha}(1,x)$ and $\Delta Q^{\alpha}(2,x)$ are discontinuities in $Q^{\alpha}(x)$ at layer 1 and 2, respectively.  $\overline{Q}^{\alpha}_x$ is the bulk, constant offset of $Q^{\alpha}_x(x)$. }
	\label{fig:chi_Q}
\end{figure}

For a partition direction ${\hat i}$, the number density operator for layer $\ell$ is
\begin{eqnarray}
n(\ell,{\hat i})_{\nu\nu'\alpha\alpha'} = \begin{cases}
\delta_{\nu\nu'}\delta_{\alpha\alpha'} ~~~~ {\rm if}~\nu \in~{\rm layer}~ \ell,\\
 0 ~~~~~~~~~~~~~ {\rm if}~\nu \notin~{\rm layer} ~\ell
\end{cases}.
\end{eqnarray}
A layer is uniquely defined by its layer number $\ell$ and orientation ${\hat i}$, and we include these arguments explicitly for relevant operators. The particle continuity equation equates the time derivative of $n(\ell,\hat{i})$ to the net influx of particle current density:
\begin{eqnarray}
\frac{d n(\ell,{\hat i})}{dt} &=&  J_i\left(\ell+\frac{1}{2}\right) - J_i\left(\ell-\frac{1}{2}\right)~.\label{eq:conteqn}
\end{eqnarray}
As described earlier, the flux density operator $J_i$ includes a subscript indicating the normal direction of the plane through which particles flow.  We use the Heisenberg equation of motion to rewrite the left-hand-side of Eq.~\ref{eq:conteqn}, obtaining the identity:
\begin{eqnarray}
\frac{1}{i\hbar}\left[H\left({\bf k}\right),n(\ell,{\hat i})\right] & = &J_i\left({\bf k},\ell+\frac{1}{2}\right) - J_i\left({\bf k},\ell-\frac{1}{2}\right).\nonumber\\ \label{eq:contEqn}
\end{eqnarray}
where we've added the Bloch wave vector ${\bf k}$ as an argument to all relevant operators (we explicitly show the ${\bf k}$-dependence of all quantities). Eq.~\ref{eq:contEqn} is often used to {\it define} the flux density operator \cite{todorov2002tight}. The terms in Eq.~\ref{eq:contEqn} and all subsequent equations correspond to matrices in the tight-binding representation.

The spin flux density operator ${\bf Q}$ is the symmetrized product of the flux density and spin operators \cite{Nagaosa_SHE,Tanaka_SHE}:
\begin{eqnarray}
Q^\alpha_i\left({\bf k},\ell+\frac{1}{2}\right) &=& \left\{ J_i\left({\bf k},\ell+\frac{1}{2}\right),s^\alpha \right\}
\end{eqnarray}
where $\{A,B\}=\frac{1}{2}(AB+BA)$. The spin flux density $Q^\alpha_i(\ell+\frac{1}{2})$ is a rank-2 tensor whose components correspond to the spin direction $\alpha$ and the direction normal to the plane through which spin flows $i$.  Applying this definition of spin flux density to Eq.~\ref{eq:contEqn}, we obtain the net spin influx on atomic layer $\ell$, which we denote as $\Delta \mathbf{Q}({\bf k},\ell,\hat{i})$:
\begin{eqnarray}
\Delta Q^{\alpha}({\bf k},\ell,\hat{i})&\equiv&Q^{\alpha}_i\left({\bf k},\ell+\frac{1}{2}\right) - Q^{\alpha}_i\left({\bf k},\ell-\frac{1}{2}\right) \nonumber\\
&=& \frac{1}{i\hbar}\left\{ \left[H({\bf k}),n(\ell,{\hat i})\right], s^{\alpha}\right\}\label{eq:scontEqn}.
\end{eqnarray}
Equation~\ref{eq:scontEqn} can be used to evaluate the position-dependent spin influx at each atomic layer. The spin influx at a layer is equal to the spin current discontinuity at each layer position.

Knowledge of the spin current discontinuity at each layer determines the ``shape'' of the spin current distribution, but does not specify the spatially constant value of the spin current.  This spatially constant component of spin current is the ``bulk'' value $\overline{\bf{Q}}$, and can be evaluated in standard fashion, as the symmetrized product of bulk velocity $dH/dk_i$ and spin operators:
	\begin{eqnarray}
	\overline{Q}^{\alpha}_i({\bf k})&=&
	\frac{1}{a_i\hbar }\left\{\frac{dH}{d{k_i}},s^\alpha \right\}.\label{eq:Qbulk}
\end{eqnarray}
In Eq.~\ref{eq:Qbulk}, $a_i$ is the unit cell length along the $\hat {i}$ direction; this factor ensures $\overline{Q}$ has units of spin current density (as opposed to a spin velocity).  As we discuss in Appendix~A, one can show that the bulk spin current equals the spatial average of the position-dependent spin current:
\begin{eqnarray}
	\overline{Q}^{\alpha}_i({\bf k})&=&\frac{1}{a_i} \sum_\ell d\left(\ell+\frac{1}{2}\right) Q_i^\alpha\left({\bf k},\ell+\frac{1}{2}\right).\label{eq:Qbulk2}
\end{eqnarray}
where $d(\ell+\frac{1}{2})$ is the distance between the adjacent layers $\ell$ and $\ell+1$.  The physical picture described by these equations is illustrated in Fig.~\ref{fig:chi_Q}:  knowledge of the average spin current, together with the discontinuities in the spin current at atomic layers, determines the full position-dependent spin current.
	
We next use this approach to compute the position-dependent intrinsic spin Hall conductivity $\sigma$.  $\sigma$ is a rank 3 tensor that relates the spin current to an applied electric field $E$: $\overline{Q}^\alpha_i(\ell+\frac{1}{2}) = \overline{\sigma}_{ji}^\alpha(\ell+\frac{1}{2}) E_j$. Note that the position-dependent $\sigma$ is defined on interlayer planes. We also evaluate the electric field-induced spin current discontinuity at each layer, denoting this response with $\chi$, so that $\Delta Q^{\alpha}(\ell,{\hat i})=\chi^{\alpha}_j(\ell,{\hat i})E_j$. $\chi$ and $\overline{\sigma}$ are evaluated with the Kubo formula expressions:
	\begin{widetext}
	\begin{eqnarray}
	\label{eq:SHC_N}
	\chi^{\alpha}_{j}(\ell,{\hat i})&=&2e^2~a_i~\text{Im}\sum_{\substack{\mathbf{k},n,\\m\neq n}}f_{n} \frac{\bra{\psi^n_{\mathbf{k}}}\frac{dH}{dk_j}\ket{\psi^m_{\mathbf{k}}}\bra{\psi^m_{\mathbf{k}}}\Delta Q^\alpha({\bf k},\ell,{\hat i})\ket{\psi^n_{\mathbf{k}}}}{(E^m_{\mathbf{k}}-E^n_{\mathbf{k}})^2+\eta^2}~~~~~~~~~\ell=\{1,2,...,N\} \\
	\overline{\sigma}^{\alpha}_{j i}&=&2e^2~a_i~\text{Im}\sum_{\substack{\mathbf{k},n,\\m\neq n}}f_n \frac{\bra{\psi^n_{\mathbf{k}}}\frac{dH}{dk_j}\ket{\psi^m_{\mathbf{k}}}\bra{\psi^m_{\mathbf{k}}}\overline{Q}^{\alpha}_i({\bf k})\ket{\psi^n_{\mathbf{k}}}}{(E^m_{\mathbf{k}}-E^n_{\mathbf{k}})^2+\eta^2}
	\end{eqnarray}
	\end{widetext}
	where $\eta$ is the smearing parameter (we use $\eta=25~\rm meV$ throughout the paper unless otherwise noted), $f_n$ is the equilibrium Fermi-Dirac distribution function, and $H_{\mathbf{k}}\ket{\psi^n_{\mathbf{k}}}=E^n_{\mathbf{k}}\ket{\psi^n_{\mathbf{k}}}$.  We adopt the prefactor $2e^2 a_i$ to ensure that the unit of $\sigma$ is conductivity $(\rm \Omega\cdot cm)^{-1} $.The $N$ unknowns $\{\sigma(\ell+\frac{1}{2})\}_{\ell=1,...,N}$ are determined by the resulting set of linear equations,
	\begin{eqnarray}
	\chi^{\alpha}_{j}(\ell,{\hat i})&=&\sigma^{\alpha}_{j i}\left(\ell+\frac{1}{2}\right)-\sigma^{\alpha}_{j i}\left(\ell-\frac{1}{2}\right)~~{\rm for}~\ell=\{1,...,N\} ~~\label{eq:sigma1}\nonumber\\ \\
	\overline{\sigma}^{\alpha}_{j i}&=&\frac{1}{a_i}\sum_\ell d\left(\ell+\frac{1}{2}\right) \sigma^\alpha_{ji}\left(\ell+\frac{1}{2}\right).\label{eq:sigma2}
	\end{eqnarray} 	
		
	The evaluation of Eqs.~\ref{eq:SHC_N}-\ref{eq:sigma2} is equally computationally intensive as the evaluation of other intrinsic response coefficients, such as the spin Hall conductivity or the spin-orbit torkance.

\subsection{Symmetry and physical considerations}
	We next discuss how the symmetry group determines the form of $\chi_{ji}^\alpha(\ell,{\hat i})$ and the ensuing form of $\sigma_{ji}^\alpha(\ell,{\hat i})$.  We represent a unitary symmetry operation $u$ with the Seitz symbol $u=\{R|t\}$ where $R$ is a local rotation operator and $t$ describes a fractional translation within the unit cell \cite{Litvin2011}. The symmetry constraint on the bulk spin Hall conductivity $\overline{\sigma}$ has been established previously \cite{Seemann2015}, and is given by:
\begin{eqnarray}
	\overline{\sigma}^{\alpha}_{ji}&=&\frac{1}{N_R}\sum_{\substack{u\in G \\l,k,\beta}} \text{det}\left(R\right)R_{lj} R_{ki} R_{\beta\alpha}~\overline{\sigma}^{\beta}_{lk}~,\label{eq:sigmaBulk}\nonumber \\
\end{eqnarray}
where the sum $u$ is over all operations in the material symmetry group $G$, and $N_R$ is the number of elements in $G$.
	
We determine the symmetry constraint on the position-resolved spin Hall conductivity $\sigma(\ell+\frac{1}{2})$ by first studying the constraints on the layer-resolved net spin influx response $\chi(\ell,{\hat i})$.  A layer $\ell$ and orientation $\hat i$ are not necessarily mapped to the same layer and orientation under a symmetry operation, so that the response on different layers may be symmetry-connected:
\begin{equation}
\chi^{\alpha}_{j}(\ell,{\hat i})=\frac{1}{N_R}\sum_{\substack{u\in G \\l,\beta}} \text{det}\left(R\right)R_{\beta\alpha}R_{lj}~\chi^{\beta}_{l}(u \ell, u{\hat i})~.\label{eq:chi1}
\end{equation}
$\chi_j^\alpha(\ell,{\hat i})$, being the net spin influx, has the additional constraint that its sum over layers in the unit cell must vanish:
\begin{eqnarray}
\sum_\ell \chi^\alpha_{j}(\ell,{\hat i}) = 0~.\label{eq:chi2}
\end{eqnarray}
This constraint can be understood by noting that the sum of individual net influxes over all atoms in the unit cell equals the total spin flux into the unit cell.  The periodic boundary condition forces the spin current at opposite sides of the unit cell to be equal, so that the total spin flux must vanish. The symmetry constraints on the position-dependent spin current response are encoded in Eqs.~\ref{eq:sigmaBulk}, \ref{eq:chi1}, and \ref{eq:chi2}.  We utilize these symmetry constraints to analyze the materials studied in the next section. 	
	
	We end this section with a brief discussion of the physics underlying this approach.  The microscopic origin of a net spin current flux into a volume is the non-conservation of spin inside that volume.  For the nonmagnetic materials studied in this work, this spin non-conservation is due to spin-orbit coupling.  Spin is one component of the system's {\it total} angular momentum, a quantity which {\it is} conserved.  In the steady state, the flux of spin angular momentum into a volume must therefore be transferred to other degrees of freedom, such as the crystal lattice \cite{haney2010current}.  See Ref.~\onlinecite{Go2020} for a derivation and discussion of the steady state angular momentum conservation equation for electrically driven systems.

	\section{Applications}
	
	We apply the method of the previous section to evaluate the position-dependent intrinsic spin Hall conductivities in \ch{WTe2} and \ch{PbTe}.  We begin with \ch{WTe2}, where we find that the layer-dependent spin Hall conductivity is consistent with the measured and computed values of the dampinglike spin-orbit torque in \ch{WTe2}-ferromagnetic bilayers.  We then consider \ch{PbTe} and find a staggered spin Hall conductivity which exceeds that of \ch{WTe2} by an order of magnitude.
	
	\subsection{WTe2}\label{Sec:WTe2}
	
\begin{figure}[htbp]
		\includegraphics[width=.95\columnwidth]{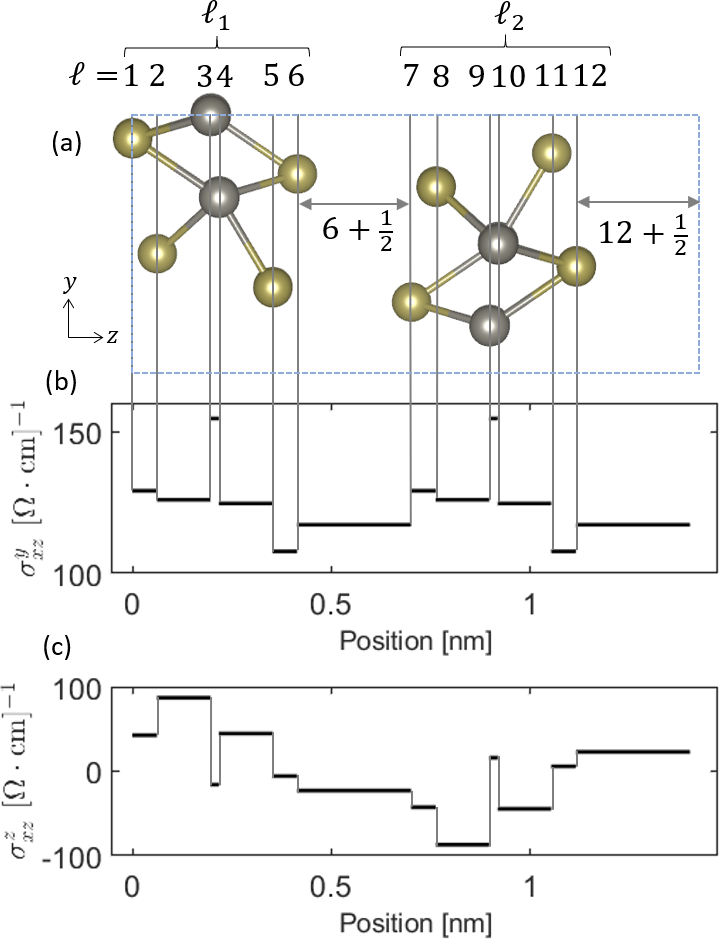}
		\caption{(a) Shows the partition of the \ch{WTe2} unit cell into layers stacked along the $\hat z$-direction. The set of layers in each monolayer of \ch{WTe2} are $\ell_1=1,...6$ and $\ell_2=7,...12$, and are connected by nonsymmorphic symmetry operations. (b) shows the position-dependent spin Hall conductivity $\sigma_{xz}^y$ and (c) shows the position-dependent $\sigma_{xz}^z$ at Fermi energy.}
		\label{Fig:WTe2lattice}
\end{figure}

	Bulk 1T'-\ch{WTe2} is orthorhombic (No. 31 $Pnm2_1$ space group) with two inequivalent layers of \ch{WTe2} \cite{Soluyanov2015,Zhou2019} along a stacking direction $z$.  We focus on flow in the $\hat z$-direction, so that the unit cell is partitioned into 12 layers, as shown in Fig.~\ref{Fig:WTe2lattice}(a).  We first analyze the symmetry constraints of the response. The symmetry operations are:
	\begin{eqnarray}
	{\rm identity}~E&:&~(x,y,z)\rightarrow(x,y,z) \nonumber\\
	{\rm mirror}~M_x&:&~(x,y,z)\rightarrow(-x,y,z)\nonumber\\
	{\rm glide~plane}~G_y&:&~(x,y,z)\rightarrow(0.5+x,-y,z+0.5)\nonumber\\
	{\rm screw}~S_z&:&~(x,y,z)\rightarrow(0.5-x,-y,z+0.5)~,\nonumber
	\end{eqnarray}
	where the translations are given in fractional coordinates.  Note the glide plane $G_y$ and screw $S_z$ symmetries are nonsymmorphic, connecting the two monolayers of \ch{WTe2}. For the bulk response, these symmetries lead to a conventional spin Hall conductivity, with mutually perpendicular electric field, flow and spin directions.

	In Appendix~\ref{App.sym}, we use these symmetry operations to explicitly evaluate Eqs.~\ref{eq:chi1} and \ref{eq:chi2}.  Layers 1-6, comprising 1 unit of \ch{WTe2}, are symmetry connected to layers 7-12, leading to a relation between the net spin influx response of the two monolayers of \ch{WTe2}:
\begin{eqnarray}
\chi_x^y\left(\ell,\hat{z}\right) &=& \chi_x^y\left(\ell+6,\hat{z}\right),\\
\chi_x^z\left(\ell,\hat{z}\right) &=& -\chi_x^z\left(\ell+6,\hat{z}\right), ~~~~~\ell=\left\{1,..,6\right\}.
\end{eqnarray}
	The above forms for $\chi$ determine the position-dependence of the spin Hall conductivity.  Evaluating Eqs.~\ref{eq:sigma1} and \ref{eq:sigma2} with the above forms for $\chi$ leads to:
	\begin{eqnarray}
	\sigma_{xz}^y(\ell+\frac{1}{2}) &=& \sigma_{xz}^y(\ell+6+\frac{1}{2}) \\
	\sigma_{xz}^z(\ell+\frac{1}{2}) &=& -\sigma_{xz}^z(\ell+6+\frac{1}{2})
	~~~~~\ell=\left\{1,..,6\right\}.
\end{eqnarray}
Fig.~\ref{Fig:WTe2lattice} (b) and (c) show that numerical calculations exactly reflect the symmetry requirement of spin Hall conductivity tensor.  Between the two \ch{WTe2} monolayers, the conventional component $\sigma_{xz}^y$ is repeated, while the unconventional component $\sigma_{xz}^z$ flips sign.

\begin{figure}[htbp]
		\includegraphics[width=1\columnwidth]{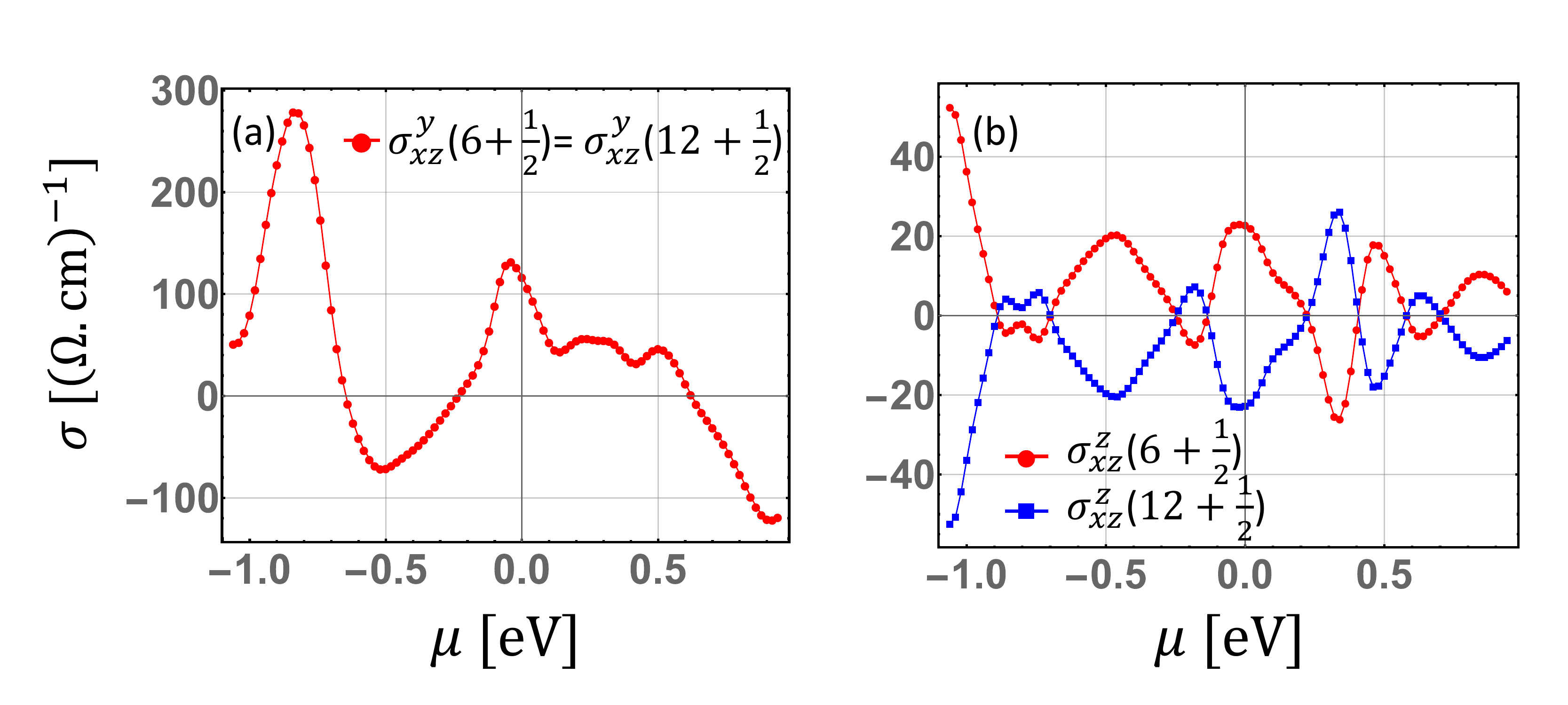}
		\caption{Spin Hall conductivity of bulk 1T'-\ch{WTe2}. (a) and (b) show the chemical potential dependence of position-dependent spin Hall conductivity tensor $\sigma^{y}_{xz}$ and $\sigma^{z}_{xz}$ respectively. Two flux plane locations are $6+\frac{1}{2}$ and $12+\frac{1}{2}$ marked in Fig.~\ref{Fig:WTe2lattice}. }
		\label{Fig:WTe2}
\end{figure}
	
Now we focus on two flux plane locations $6+\frac{1}{2}$ and $12+\frac{1}{2}$ which separate two monolayers of \ch{WTe2}. Fig.~\ref{Fig:WTe2} shows the first-principles results of the two nonzero spin current conductivity components as a function of chemical potential. $\sigma_{xz}^y$ is the conventional spin Hall effect in which the electric field direction, spin current flow direction, and spin polarization are all perpendicular to each other. $\sigma_{xz}^z$ is the unconventional component in which spin polarization and spin current flow direction are parallel. Although the total $\sigma^z_{xz}$ in bulk $\ch{WTe2}$ is zero, the magnitude of the spin current conductivity between each individual layer is around 20 $(\rm \Omega\cdot cm)^{-1}$ near the Fermi level. Note that the conventional spin Hall component $\sigma^y_{xz}$ has larger magnitude, around 100 $(\rm \Omega\cdot cm)^{-1}$ \cite{Zhou2019,Dash2020Conventional}, near the Fermi level.
	
The $\sigma^y_{xz}$ and $\sigma^z_{xz}$ components of the spin Hall conductivity are of particular interest in light of recent experiments involving heterostructures composed of \ch{WTe2} and thin film ferromagnets \cite{MacNeill2016,MacNeill2017,shi2019all}.  The $\sigma^y_{xz}$ component contributes to the conventional dampinglike torque while the $\sigma^z_{xz}$ component contributes to the unconventional dampinglike torque which drives the magnetization to an out-of-plane configuration \cite{MacNeill2016,MacNeill2017,Xue2020SOT}. We find that the magnitude of both uniform and staggered spin current conductivity tensor is similar to the magnitude of spin-orbit torque in experiments \cite{MacNeill2016} and calculations \cite{Xue2020SOT}. This observation indicates that the staggered spin Hall effect can explain the unconventional spin-orbit torque exerted on an adjacent ferromagnet. The calculation of the bulk \ch{WTe2} staggered spin Hall conductivity provides a computationally much cheaper estimation of unconventional component of the spin-orbit torque as compared to a direct computation of the torque in a heterostructure.

	\subsection{\ch{PbTe}}

	\begin{figure}[htbp]
		\includegraphics[width=1\columnwidth]{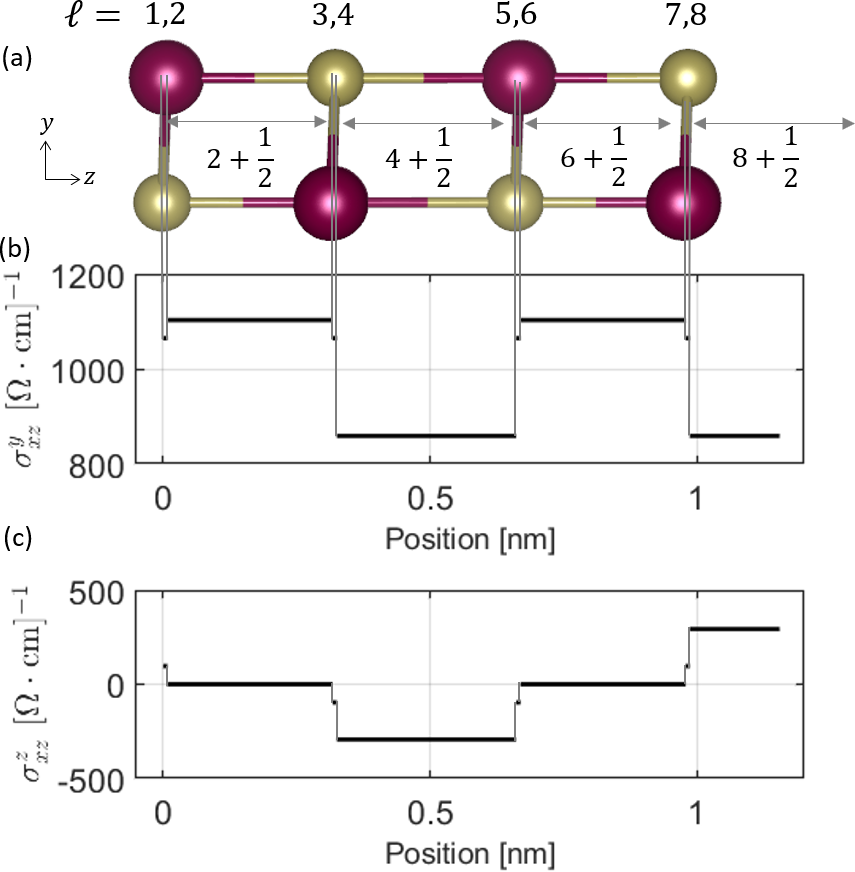}
		\caption{(a) Shows the partition of the PbTe unit cell into layers stacked along the $\hat z$-direction.  (b) shows the position-dependent spin Hall conductivity $\sigma_{xz}^y$ and (c) shows the position-dependent $\sigma_{xz}^z$ at chemical potential $\mu=-0.38~\rm eV$.}
		\label{Fig:PbTelattice}
	\end{figure}
	\begin{figure}[htbp]
		\includegraphics[width=1\columnwidth]{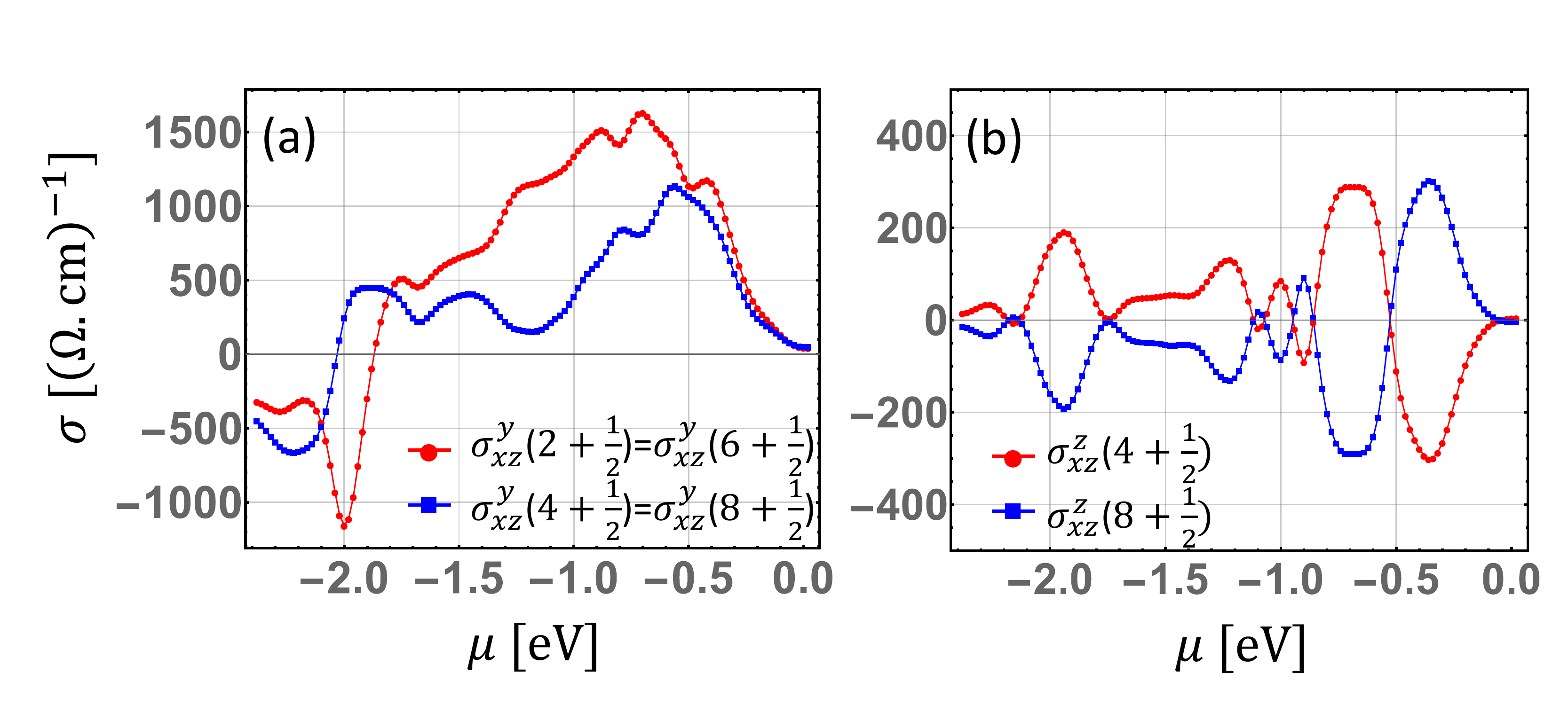}
		\caption{Spin Hall conductivity of orthorhomic phase \ch{PbTe}. (a) and (b) show the chemical potential dependence of position-dependent spin Hall conductivity tensor $\sigma^{y}_{xz}$ and $\sigma^{z}_{xz}$ respectively. Four flux plane locations are marked in Fig.~\ref{Fig:PbTelattice}(a). In (b),  we do not plot $\sigma^{z}_{xz}(2+\frac{1}{2})$ and $\sigma^{z}_{xz}(6+\frac{1}{2})$ because they are exactly zero.}
		\label{Fig:PbTe}
	\end{figure}
	
We next consider \ch{PbTe}, a material which is less well studied for spintronics applications.  \ch{PbTe} belongs to the family of narrow band gap IV-VI semiconductors which transforms from the rock salt structure to the orthorhombic $Pnma$ (No. 62) phase under high pressure \cite{PbTe1967}.
Although the $Pnma$ phase of \ch{PbTe} is not the ground state under ambient conditions, it serves as an example of a relatively large hidden spin Hall conductivity. The $Pnma$ phase of \ch{PbTe} has 4 Pb atoms and 4 Te atoms in the orthorhombic unit cell, as shown in Fig.~\ref{Fig:PbTelattice}(a). We again focus on flow in the $\hat z$-direction, and partition the unit cell into 8 layers, shown in Fig.~\ref{Fig:PbTelattice}(a).

Without translation in the $z$-direction, each individual layer only preserves one mirror symmetry $(x\rightarrow-x)$, similar to 1T'-\ch{WTe2}. With additional two glide reflections (in the $xz$ and $xy$ planes) and an inversion center, the unconventional component $\sigma^z_{xz}$ vanishes in the bulk and only the conventional component $\sigma^y_{xz}$ survives.  A symmetry analysis yields the following relations between the position-dependent spin Hall conductivity components (see App.~\ref{App.sym}):
\begin{eqnarray}
\label{eq:PbTesym}
	\sigma^y_{xz}(\ell+\frac{1}{2})&=&\sigma^y_{xz}(\ell+4+\frac{1}{2}), \nonumber\\
	\sigma^z_{xz}(\ell+\frac{1}{2})&=&-\sigma^z_{xz}(\ell+4+\frac{1}{2}),~~~~~\ell=\left\{1,..,4\right\}.
\end{eqnarray}
Figure~\ref{Fig:PbTelattice}(b) and (c) show our numerical results of position-dependent spin Hall conductivity tensor $\sigma^y_{xz}$ and $\sigma^z_{xz}$.  Note that symmetry dictates that the unconventional component, with spin and flow direction along $z$, vanishes at planes $2+\frac{1}{2}$ and $6+\frac{1}{2}$: $\sigma^z_{xz}(2+\frac{1}{2})=\sigma^z_{xz}(6+\frac{1}{2})=0$. Additionally, there is no symmetry-derived relation between the conventional component of spin current passing though planes $2+\frac{1}{2}~(6+\frac{1}{2})$ and planes $4+\frac{1}{2}~(8+\frac{1}{2})$, so that this component of the spin current generally varies in space.
	
Figure~\ref{Fig:PbTe} summarizes our calculations of the spin Hall conductivity components $\sigma^y_{xz}$ and $\sigma^z_{xz}$ as a function of chemical potential. We focus on the hole-doped case where the chemical potential is measured with respect to the valence band maximum. In Fig.~\ref{Fig:PbTe}a we find that conventional spin Hall conductivity $\sigma^y_{xz}$ can be quite large, more than $1000~ ~(\rm \Omega\cdot cm)^{-1}$, at some chemical potentials. Additionally, we find the symmetry-required relations  $\sigma^y_{xz}(2+\frac{1}{2})=\sigma^y_{xz}(6+\frac{1}{2})$ and $\sigma^y_{xz}(4+\frac{1}{2})=\sigma^y_{xz}(8+\frac{1}{2})$ are satisfied.

The numerical results of \ch{PbTe} indicate that these doped orthorhombic IV-VI semiconductors are candidates for the efficient generation of both conventional $Q^y_z$ and unconventional $Q^z_z$ spin current flowing along ${\hat z}$-direction with applied electric filed in ${\hat x}$-direction. This real-space staggered spin current generation can be used to exert a torque on the adjacent perpendicularly-magnetized ferromagnets with similar experimental setups of 1T'-\ch{WTe2}/ ferromagnets \cite{MacNeill2016}.

\section{Conclusion} In this paper we present a method to compute the position-dependent intrinsic spin Hall conductivity.  This method requires similar computational effort as other standard calculations of instrinsic response, {\it e.g.}, anomalous Hall effect.  We verify the validity of this method by direct comparison to other, more direct approaches to computing the position-dependent spin current. Note that our method focuses on how the electric field-induced spin current varies within a \textit{periodic} unit cell; this is more narrow in scope than recent work \cite{Tomas2020} which derives the general local form for the spin Hall conductivity.  We also derive the symmetry constrained form of the position-dependent spin current response.  This general expression is useful for efficiently determining a class of materials which exhibit a particular desired response.  We apply this method to compute the position-dependent spin Hall conductivity for \ch{WTe2} and \ch{PbTe}. For \ch{WTe2}, we find the bulk conventional and staggered unconventional spin Hall conductivity values are similar to the torque measured and computed in \ch{WTe2}-ferromagnet bilayers.  This indicates that, at least for this material, the torque in a bilayer heterostructure can be estimated {\it a priori} by computing the bulk spin Hall conductivity, which is computationally much less intensive than directly computing the torque in a heterostructure.  We next compute the response for \ch{PbTe}, and find an order of magnitude large response relative to \ch{WTe2}, indicating the potential usefulness of this material heterostructures with a ferromagnet.

	\section{Acknowledgment}
	We thank Mark D. Stiles and Emilie Ju\'{e} for carefully reading the manuscript and providing insightful
comments. F.X. acknowledges support under the Cooperative Research Agreement between the University of Maryland and the National Institute of Standards and Technology Physical Measurement Laboratory, Award 70NANB14H209, through the University of Maryland.

	\appendix
	
	\section{Direction construction of position-dependent spin current operator}
	\label{App.position}

	Here we describe the method for directly computing the position-dependent spin current operators. It suffices to consider a 1-d system, as shown in Fig. \ref{fig:lattice}.  Below we show the Hamiltonian for an infinite chain with 2nd nearest-neighbor hopping:
	\begin{equation}
	H = \begin{pmatrix}
	\ddots & \vdots & \vdots & \vdots & \vdots& \vdots  \\
	\dots & U & t  & t'  & 0 & 0 & \dots \\
	\dots  & t^\dagger  & U & t  & t' & 0 & \dots\\
	\dots  & (t')^\dagger  & t^\dagger  & U & t & t' & \dots\\
	\dots & 0 & (t')^\dagger & t^\dagger & U & t& \dots\\
	\dots & 0 & 0 & (t')^\dagger & t\dagger & U & \dots \\
	& \vdots & \vdots & \vdots & \vdots & \vdots & \ddots \\
	\end{pmatrix}.
	\label{eq:Hchain}
	\end{equation}
	$U$ is a matrix describing the primary unit cell.  $t$ ($t'$) is a matrix describing the coupling, or ``hopping'', between a primary unit cell and its copy, translated by 1 (2) lattice spacings. Generally, one invokes Bloch's theorem to map this infinite system to the primary unit cell:
	\begin{eqnarray}
	H_0(k)\phi_0=E\phi_0~,
	\end{eqnarray}
	where $H_0(k) = (U/2 + t e^{ika} + t' e^{2ika}) + {\rm h.c.}$.  From this, the eigenvector $\phi_0$ and eigenvalue $E$ are readily obtained.  The current averaged over the primary unit cell is obtained with the operator $dH_0/dk$.
	
	To construct the position-dependent current operator, the infinite matrix Eq.~\ref{eq:Hchain} with $N$ nearest neighbors is first truncated to dimension $\left(2N+1\right)\times \left(2N+1\right)$. In our example, $N=2$ (2nd nearest-neighbor hopping) so that we use a $5\times 5$ matrix:
	\begin{equation}
	H^s= \begin{pmatrix}
	U & t & t' & 0 & 0  \\
	t^\dagger  & U & t  & t'  & 0\\
	(t')^\dagger & t^\dagger  & U & t  & t'\\
	0 & (t')^\dagger & t^\dagger  & U & t\\
	0 & 0 & (t')^\dagger & t^\dagger & U \\
	\end{pmatrix},
	\label{eq:Hfinite}
	\end{equation}
	The resulting finite matrix equation satisfies Schrodinger equation only in the primary unit cell at the center of the finite system (for this case, site 3).
	\begin{eqnarray}
	H^s \begin{pmatrix}
	\phi_0 e^{-2ika}\\
	\phi_0 e^{-ika}\\
	\phi_0\\
	\phi_0 e^{+ika}\\
	\phi_0 e^{+2ika}
	\end{pmatrix}
 = \begin{pmatrix}
	-\\
	-\\
	E \phi_0\\
	-\\
	-
	\end{pmatrix}
	\end{eqnarray}
	The blank components of the vector on the right-hand-side of the above equation are unknown, but are not important.  We can utilize the validity of the Schrodinger equation in the central unit cell to construct position-dependent current operators in that region.  In our example. the current operators $J_2$ and $J_3$ (depicted in Fig.~\ref{fig:lattice}) are given as:
	
	\begin{figure}
		\includegraphics[width=1.0\columnwidth]{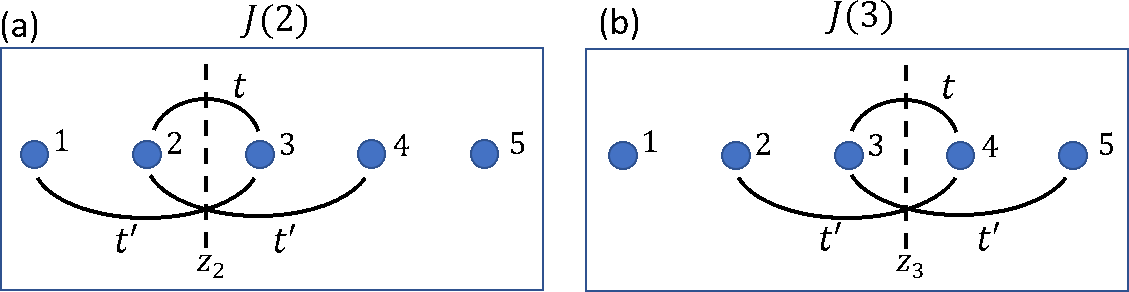}
		\caption{(a) Schematic of interactions entering the flux operator for $z=z_2$.  (b) Same schematic for $z=z_3$.} \label{fig:lattice}
	\end{figure}
	
	\begin{eqnarray}
	J_2 &=&i\left(\begin{array}{ccccc}
	0 &  0  &  t' & 0  & 0  \\
	0 &  0  &  t   & t' & 0 \\
	-(t')^\dagger &  -t^\dagger  &  0  & 0  & 0 \\
	0 &  -(t')^\dagger  &  0  & 0  & 0 \\
	0 &  0  &  0  & 0  & 0  \\
	\end{array}\right)\nonumber \\
	J_3 &=&i\left(\begin{array}{ccccc}
	0 &  0  &  0 & 0  & 0  \\
	0 &  0  &  0  & t'  & 0 \\
	0 &  0  &  0  & t   & t'  \\
	0 &  -(t')^\dagger &  -t^\dagger  & 0  & 0 \\
	0 &  0  &  -(t')^\dagger  & 0  & 0  \\
	\end{array}\right)\nonumber
	\end{eqnarray}
	The position-dependent current for state $\phi_0$ is evaluated using $\phi^s = (\phi_0 e^{-2ika},\phi_0 e^{-ika},\phi_0,\phi_0, e^{ika}\phi_0 e^{2ika})^T$ and the operators above: $\langle J_{2,3}\rangle = \langle \phi^s | J_{2,3} | \phi^s \rangle$.
	
	In general, the current operator for a plane $z_0$ is:
	\begin{eqnarray}
	J_{z_0} = \sum_{\substack{j \in z_L\\k \in z_R}} i \left(H_{jk} c_j^\dagger c_k - H_{kj} c_k^\dagger c_j\right)
	\end{eqnarray}
	where $z_L$ is the set of sites with $z$ coordinate less than $z_0$, and $z_R$ is the set of sites with $z$ coordinate greater than $z_0$.  We emphasize that with this approach, $z_0$ must lie within the central unit cell (site 3 in this example). We have checked that the position-dependent spin current obtained with this direct construction of the spin current operator agrees ``exactly'' (to within machine precision) with the alternative approach given in the main text.
	
	The spin Hall conductivity is a piecewise constant function having finite number of pieces depending on the atomic locations shown in Fig~\ref{Fig:WTe2lattice} and Fig~\ref{Fig:PbTelattice}. 	With the explicit form of the position-dependent current operators, one can prove the equality Eq.~\ref{eq:Qbulk2} of the main text by directly evaluating both sides of the equation: the ``integration'' of position-dependent spin Hall conductivity is identical to the bulk spin Hall conductivity.
	
	This direct method for computing the spin current is less desirable than the approach described in the main text because it requires the additional overhead of constructing the current operator matrices which can be quite large depending on the hopping cutoff in the constructed supercell.

	\section{First principles details}
	
	We provide technical details for the first principles calculations presented in the main text for 1T'-\ch{WTe2} and orthorhombic \ch{PbTe}. We use Quantum ESPRESSO \cite{QE} and Wannier90 \cite{Wannier90} to obtain the spin-orbit coupled localized orbital Hamiltonian in the atomic basis. In the Quantum ESPRESSO implementation, we use the pseudopotentials from PSlibrary \cite{DALCORSO2014337} generated with a fully relativistic calculation using Projector Augmented-Wave method \cite{PAW} and Perdew-Burke-Ernzerhof exchange correlations \cite{PBEGGA}. We utilize a $12\times 10 \times 6~(11\times 12 \times 4)$ Monkhorst-Pack mesh \cite{MPmesh}, $1088~{\rm eV}$ cutoff energy for \ch{WTe2} (\ch{PbTe}). We project plane-wave solutions onto atomic $d$ orbitals of transition metal atoms, $s$ and $p$ orbitals of chalcogen atoms and Pb atoms. We then symmetrize the spin-orbit coupled Wannier Hamiltonian \cite{TBmodels} since the presence of slight asymmetry could result non-vanishing staggered spin Hall conductivity. We perform a dense $k$ mesh of $480\times270\times120$ ($480\times490\times170$) to evaluate the spin current conductivity for \ch{WTe2} (\ch{PbTe}). In the implementation of Eq.~\ref{eq:SHC_N}, we adopt the approximation \cite{Souza2019} that Wannier orbitals are perfectly localized on atomic sites and spin matrix is half of Pauli matrix in the space spanned by Wannier orbitals.
	
	\section{Symmetry Analysis}\label{App.sym}	
	In this section we provide the explicit evaluation of Eqs.~\ref{eq:chi1} and \ref{eq:chi2} for \ch{WTe2} and \ch{PbTe}.  
	\subsection{\ch{WTe2}}
	The layers of the unit cell stacked along the $\hat z$-direction are shown in Fig.~\ref{Fig:WTe2lattice}(a).  We first introduce notation to distinguish the two halves of the \ch{WTe2} atoms comprising the unit cell.  We denote the set of layers 1-6 with $\ell_1$, and the set of layers 7-12 with $\ell_2$.  $\ell_1$ and $\ell_2$ are related via: $\ell_2=\ell_1+6$. We start with the most general form of the net spin current flux tensor $\chi$ at layer $\ell$:
\begin{eqnarray}
	\chi\left(\ell,{\hat z}\right) &=& \begin{pmatrix}
	\chi^x_{x}(\ell,{\hat z}) & \chi^x_{y}(\ell,{\hat z}) & \chi^x_{z}(\ell,{\hat z})\\
	\chi^y_{x}(\ell,{\hat z}) & \chi^y_{y}(\ell,{\hat z}) & \chi^y_{z}(\ell,{\hat z})  \\
	\chi^z_{x}(\ell,{\hat z}) & \chi^z_{y}(\ell,{\hat z}) & \chi^z_{z}(\ell,{\hat z})
	\end{pmatrix}.
\end{eqnarray}

The representation of the rotation parts of the symmetry operations are given by:
\begin{eqnarray}
	M_x&=&\begin{pmatrix}
	-1 & 0 & 0\\
	0 & 1 & 0  \\
	0 & 0 & 1
	\end{pmatrix}\\
	G_y&=&\begin{pmatrix}
		1 & 0 & 0\\
		0 & -1 & 0  \\
		0 & 0 & 1
		\end{pmatrix}\\
		S_z&=&\begin{pmatrix}
		-1 & 0 & 0\\
		0 & -1 & 0  \\
		0 & 0 & 1
		\end{pmatrix}.
\end{eqnarray}
We note again that the symmetry operations $G_y$ and $S_z$ involve translation along the $z$-direction, mapping layer $\ell_1$ to layer $\ell_2$.
	
We first consider how the symmorphic symmetry $M_x$ constrains the response by applying Eq.~\ref{eq:chi1} with only this operation:
\begin{eqnarray}
	\chi\left(\ell,{\hat z}\right)=\frac{1}{2}\left(\chi\left( \ell,{\hat z}\right)-M_x^T\chi\left(\ell,{\hat z}\right)M_x\right),
\end{eqnarray}
where the minus sign comes from ${\rm det}(M_x)$. The resulting $\chi\left(\ell,{\hat z}\right)$ has the form:
	\begin{eqnarray}
	\chi\left(\ell,{\hat z}\right) &=& \begin{pmatrix}
	0 & \chi^x_{y}(\ell,{\hat z}) & \chi^x_{z}(\ell,{\hat z})\\
	\chi^y_{x}(\ell,{\hat z}) & 0 & 0  \\
	\chi^z_{x}(\ell,{\hat z}) & 0 & 0
	\end{pmatrix},
	\end{eqnarray}	
which allows the unconventional component $\chi^z_x$. Now we apply both symmorphic and nonsymmorphic symmetries using Eq.~\ref{eq:chi2},
		\begin{eqnarray}
		\chi\left(\ell_1,{\hat z}\right)&=&\frac{1}{4}\left(\chi\left(\ell_1,{\hat z}\right)-M_x^{T}\chi\left(\ell_1,{\hat z}\right)M_x \right. \nonumber \\ && ~~~ \left. -G_y^{T}\chi\left(\ell_2,{\hat z}\right)G_y+S_z^{T}\chi\left(\ell_2,{\hat z}\right)S_z\right)\\
		\chi\left(\ell_2,{\hat z}\right)&=&\frac{1}{4}\left(\chi\left(\ell_2,{\hat z}\right)-M_x^{T}\chi\left(\ell_2,{\hat z}\right)M_x \right. \nonumber \\ && ~~~ \left .-G_y^{T}\chi\left(\ell_1,{\hat z}\right)G_y+S_z^{T}\chi\left(\ell_1,{\hat z}\right)S_z\right).
		\end{eqnarray}
The resulting tensors are
		\begin{widetext}
		\begin{eqnarray}
		\chi\left(\ell_1,{\hat z}\right) &=& \begin{pmatrix}
		0 & \frac{1}{2}\big(\chi^x_{y}(\ell_1,{\hat z})+\chi^x_{y}(\ell_2,{\hat z})\big) & \frac{1}{2}\big(\chi^x_{z}(\ell_1,{\hat z})-\chi^x_{z}(\ell_2,{\hat z})\big)\\
		\frac{1}{2}\big(\chi^y_{x}(\ell_1,{\hat z})+\chi^y_{x}(\ell_2,{\hat z})\big) & 0 & 0  \\
		\frac{1}{2}\big(\chi^z_{x}(\ell_1,{\hat z})-\chi^z_{x}(\ell_2,{\hat z})\big) & 0 & 0
		\end{pmatrix},\\
	\chi\left(\ell_2,{\hat z}\right) &=& \begin{pmatrix}
		0 & \frac{1}{2}\big(\chi^x_{y}(\ell_1,{\hat z})+\chi^x_{y}(\ell_2,{\hat z})\big) & \frac{1}{2}\big(\chi^x_{z}(\ell_2,{\hat z})-\chi^x_{z}(\ell_1,{\hat z})\big)\\
		\frac{1}{2}\big(\chi^y_{x}(\ell_1,{\hat z})+\chi^y_{x}(\ell_2,{\hat z})\big) & 0 & 0  \\
		\frac{1}{2}\big(\chi^z_{x}(\ell_2,{\hat z})-\chi^z_{x}(\ell_1,{\hat z})\big) & 0 & 0
		\end{pmatrix}.
		\end{eqnarray}	

	The remaining constraint is that $\sum_\ell \chi(\ell,{\hat z})=0$, as discussed in the main text.  This implies that the $xy$ and $yx$ components of each tensor above vanishes: $\chi_y^x(\ell_1)+\chi_y^x(\ell_2)=0$, while it imposes no additional constraint on the $xz$ and $zx$ components.  One finally obtains:

	\begin{eqnarray}
	\chi\left(\ell_1\right) &=& \begin{pmatrix}
		0 & 0 & \frac{1}{2}\big(\chi^x_{z}(\ell_1,{\hat z})-\chi^x_{z}(\ell_2,{\hat z})\big)\\
		0 & 0 & 0  \\
		\frac{1}{2}\big(\chi^z_{x}(\ell_1,{\hat z})-\chi^z_{x}(\ell_2,{\hat z})\big) & 0 & 0
		\end{pmatrix}\nonumber \\
		&=& - 	\chi\left(\ell_2,{\hat z}\right)
	\end{eqnarray}

\subsection{\ch{PbTe}}
A similar symmetry analysis can be performed for \ch{PbTe}. After summing over symmetry operations, the $\chi$ tensors at four different layers, $\chi(\ell_1)=\chi(1)+\chi(2),\chi(\ell_2)=\chi(3)+\chi(4), \chi(\ell_3)=\chi(5)+\chi(6),\chi(\ell_4)=\chi(7)+\chi(8)$, are
	\begin{eqnarray}
	\chi\left(\ell_1\right) &=& \begin{pmatrix}
	0 & \chi^x_{y} & \chi^x_{z}\\
	\chi^y_{x} & 0 & 0 \\
	\chi^z_{x} & 0 & 0
	\end{pmatrix}. \\
	\chi\left(\ell_2\right) &=& \begin{pmatrix}
	0 & -\chi^x_{y} & \chi^x_{z}\\
	-\chi^y_{x} & 0 & 0 \\
	\chi^z_{x} & 0 & 0
	\end{pmatrix}.\\
	\chi\left(\ell_3\right) &=& \begin{pmatrix}
	0 & \chi^x_{y} & -\chi^x_{z}\\
	\chi^y_{x} & 0 & 0 \\
	-\chi^z_{x} & 0 & 0
	\end{pmatrix}.\\
	\chi\left(\ell_4\right) &=& \begin{pmatrix}
	0 & -\chi^x_{y} & -\chi^x_{z}\\
	-\chi^y_{x} & 0 & 0 \\
	-\chi^z_{x} & 0 & 0
	\end{pmatrix}.
	\end{eqnarray}
Note that these tensors automatically satisfy $\sum_\ell \chi(\ell)=0$, so that this condition adds no new constraints.  However, the global system symmetry requires $\sigma_{xz}^z=0$, so that $\sum_\ell \sigma^{z}_{xz}(\ell+\frac{1}{2}) =0$.  From this, we obtain the relations between position-dependent spin conductivity components, given by Eq.~\ref{eq:PbTesym} of the main text.
	\end{widetext}

	\bibliography{reference}{}

\providecommand{\noopsort}[1]{}\providecommand{\singleletter}[1]{#1}%
\begin{thebibliography}{41}%
\makeatletter
\providecommand \@ifxundefined [1]{%
 \@ifx{#1\undefined}
}%
\providecommand \@ifnum [1]{%
 \ifnum #1\expandafter \@firstoftwo
 \else \expandafter \@secondoftwo
 \fi
}%
\providecommand \@ifx [1]{%
 \ifx #1\expandafter \@firstoftwo
 \else \expandafter \@secondoftwo
 \fi
}%
\providecommand \natexlab [1]{#1}%
\providecommand \enquote  [1]{``#1''}%
\providecommand \bibnamefont  [1]{#1}%
\providecommand \bibfnamefont [1]{#1}%
\providecommand \citenamefont [1]{#1}%
\providecommand \href@noop [0]{\@secondoftwo}%
\providecommand \href [0]{\begingroup \@sanitize@url \@href}%
\providecommand \@href[1]{\@@startlink{#1}\@@href}%
\providecommand \@@href[1]{\endgroup#1\@@endlink}%
\providecommand \@sanitize@url [0]{\catcode `\\12\catcode `\$12\catcode
  `\&12\catcode `\#12\catcode `\^12\catcode `\_12\catcode `\%12\relax}%
\providecommand \@@startlink[1]{}%
\providecommand \@@endlink[0]{}%
\providecommand \url  [0]{\begingroup\@sanitize@url \@url }%
\providecommand \@url [1]{\endgroup\@href {#1}{\urlprefix }}%
\providecommand \urlprefix  [0]{URL }%
\providecommand \Eprint [0]{\href }%
\providecommand \doibase [0]{https://doi.org/}%
\providecommand \selectlanguage [0]{\@gobble}%
\providecommand \bibinfo  [0]{\@secondoftwo}%
\providecommand \bibfield  [0]{\@secondoftwo}%
\providecommand \translation [1]{[#1]}%
\providecommand \BibitemOpen [0]{}%
\providecommand \bibitemStop [0]{}%
\providecommand \bibitemNoStop [0]{.\EOS\space}%
\providecommand \EOS [0]{\spacefactor3000\relax}%
\providecommand \BibitemShut  [1]{\csname bibitem#1\endcsname}%
\let\auto@bib@innerbib\@empty
\bibitem [{\citenamefont {Sinova}\ \emph {et~al.}(2015)\citenamefont {Sinova},
  \citenamefont {Valenzuela}, \citenamefont {Wunderlich}, \citenamefont
  {Back},\ and\ \citenamefont {Jungwirth}}]{SinovaRMP2015}%
  \BibitemOpen
  \bibfield  {author} {\bibinfo {author} {\bibfnamefont {J.}~\bibnamefont
  {Sinova}}, \bibinfo {author} {\bibfnamefont {S.~O.}\ \bibnamefont
  {Valenzuela}}, \bibinfo {author} {\bibfnamefont {J.}~\bibnamefont
  {Wunderlich}}, \bibinfo {author} {\bibfnamefont {C.~H.}\ \bibnamefont
  {Back}},\ and\ \bibinfo {author} {\bibfnamefont {T.}~\bibnamefont
  {Jungwirth}},\ }\href {https://doi.org/10.1103/RevModPhys.87.1213} {\bibfield
   {journal} {\bibinfo  {journal} {Rev. Mod. Phys.}\ }\textbf {\bibinfo
  {volume} {87}},\ \bibinfo {pages} {1213} (\bibinfo {year}
  {2015})}\BibitemShut {NoStop}%
\bibitem [{\citenamefont {Hirsch}(1999)}]{Hirsch1999}%
  \BibitemOpen
  \bibfield  {author} {\bibinfo {author} {\bibfnamefont {J.~E.}\ \bibnamefont
  {Hirsch}},\ }\href {https://doi.org/10.1103/PhysRevLett.83.1834} {\bibfield
  {journal} {\bibinfo  {journal} {Phys. Rev. Lett.}\ }\textbf {\bibinfo
  {volume} {83}},\ \bibinfo {pages} {1834} (\bibinfo {year}
  {1999})}\BibitemShut {NoStop}%
\bibitem [{\citenamefont {Sinova}\ \emph {et~al.}(2004)\citenamefont {Sinova},
  \citenamefont {Culcer}, \citenamefont {Niu}, \citenamefont {Sinitsyn},
  \citenamefont {Jungwirth},\ and\ \citenamefont {MacDonald}}]{Sinova2004}%
  \BibitemOpen
  \bibfield  {author} {\bibinfo {author} {\bibfnamefont {J.}~\bibnamefont
  {Sinova}}, \bibinfo {author} {\bibfnamefont {D.}~\bibnamefont {Culcer}},
  \bibinfo {author} {\bibfnamefont {Q.}~\bibnamefont {Niu}}, \bibinfo {author}
  {\bibfnamefont {N.~A.}\ \bibnamefont {Sinitsyn}}, \bibinfo {author}
  {\bibfnamefont {T.}~\bibnamefont {Jungwirth}},\ and\ \bibinfo {author}
  {\bibfnamefont {A.~H.}\ \bibnamefont {MacDonald}},\ }\href
  {https://doi.org/10.1103/PhysRevLett.92.126603} {\bibfield  {journal}
  {\bibinfo  {journal} {Phys. Rev. Lett.}\ }\textbf {\bibinfo {volume} {92}},\
  \bibinfo {pages} {126603} (\bibinfo {year} {2004})}\BibitemShut {NoStop}%
\bibitem [{\citenamefont {Liu}\ \emph {et~al.}(2011)\citenamefont {Liu},
  \citenamefont {Moriyama}, \citenamefont {Ralph},\ and\ \citenamefont
  {Buhrman}}]{liu2011spin}%
  \BibitemOpen
  \bibfield  {author} {\bibinfo {author} {\bibfnamefont {L.}~\bibnamefont
  {Liu}}, \bibinfo {author} {\bibfnamefont {T.}~\bibnamefont {Moriyama}},
  \bibinfo {author} {\bibfnamefont {D.}~\bibnamefont {Ralph}},\ and\ \bibinfo
  {author} {\bibfnamefont {R.}~\bibnamefont {Buhrman}},\ }\href@noop {}
  {\bibfield  {journal} {\bibinfo  {journal} {Physical review letters}\
  }\textbf {\bibinfo {volume} {106}},\ \bibinfo {pages} {036601} (\bibinfo
  {year} {2011})}\BibitemShut {NoStop}%
\bibitem [{\citenamefont {Miron}\ \emph {et~al.}(2011)\citenamefont {Miron},
  \citenamefont {Garello}, \citenamefont {Gaudin}, \citenamefont {Zermatten},
  \citenamefont {Costache}, \citenamefont {Auffret}, \citenamefont {Bandiera},
  \citenamefont {Rodmacq}, \citenamefont {Schuhl},\ and\ \citenamefont
  {Gambardella}}]{miron2011perpendicular}%
  \BibitemOpen
  \bibfield  {author} {\bibinfo {author} {\bibfnamefont {I.~M.}\ \bibnamefont
  {Miron}}, \bibinfo {author} {\bibfnamefont {K.}~\bibnamefont {Garello}},
  \bibinfo {author} {\bibfnamefont {G.}~\bibnamefont {Gaudin}}, \bibinfo
  {author} {\bibfnamefont {P.-J.}\ \bibnamefont {Zermatten}}, \bibinfo {author}
  {\bibfnamefont {M.~V.}\ \bibnamefont {Costache}}, \bibinfo {author}
  {\bibfnamefont {S.}~\bibnamefont {Auffret}}, \bibinfo {author} {\bibfnamefont
  {S.}~\bibnamefont {Bandiera}}, \bibinfo {author} {\bibfnamefont
  {B.}~\bibnamefont {Rodmacq}}, \bibinfo {author} {\bibfnamefont
  {A.}~\bibnamefont {Schuhl}},\ and\ \bibinfo {author} {\bibfnamefont
  {P.}~\bibnamefont {Gambardella}},\ }\href@noop {} {\bibfield  {journal}
  {\bibinfo  {journal} {Nature}\ }\textbf {\bibinfo {volume} {476}},\ \bibinfo
  {pages} {189} (\bibinfo {year} {2011})}\BibitemShut {NoStop}%
\bibitem [{\citenamefont {Liu}\ \emph {et~al.}(2012{\natexlab{a}})\citenamefont
  {Liu}, \citenamefont {Pai}, \citenamefont {Li}, \citenamefont {Tseng},
  \citenamefont {Ralph},\ and\ \citenamefont {Buhrman}}]{liu2012spin}%
  \BibitemOpen
  \bibfield  {author} {\bibinfo {author} {\bibfnamefont {L.}~\bibnamefont
  {Liu}}, \bibinfo {author} {\bibfnamefont {C.-F.}\ \bibnamefont {Pai}},
  \bibinfo {author} {\bibfnamefont {Y.}~\bibnamefont {Li}}, \bibinfo {author}
  {\bibfnamefont {H.}~\bibnamefont {Tseng}}, \bibinfo {author} {\bibfnamefont
  {D.}~\bibnamefont {Ralph}},\ and\ \bibinfo {author} {\bibfnamefont
  {R.}~\bibnamefont {Buhrman}},\ }\href@noop {} {\bibfield  {journal} {\bibinfo
   {journal} {Science}\ }\textbf {\bibinfo {volume} {336}},\ \bibinfo {pages}
  {555} (\bibinfo {year} {2012}{\natexlab{a}})}\BibitemShut {NoStop}%
\bibitem [{\citenamefont {Liu}\ \emph {et~al.}(2012{\natexlab{b}})\citenamefont
  {Liu}, \citenamefont {Lee}, \citenamefont {Gudmundsen}, \citenamefont
  {Ralph},\ and\ \citenamefont {Buhrman}}]{liu2012current}%
  \BibitemOpen
  \bibfield  {author} {\bibinfo {author} {\bibfnamefont {L.}~\bibnamefont
  {Liu}}, \bibinfo {author} {\bibfnamefont {O.}~\bibnamefont {Lee}}, \bibinfo
  {author} {\bibfnamefont {T.}~\bibnamefont {Gudmundsen}}, \bibinfo {author}
  {\bibfnamefont {D.}~\bibnamefont {Ralph}},\ and\ \bibinfo {author}
  {\bibfnamefont {R.}~\bibnamefont {Buhrman}},\ }\href@noop {} {\bibfield
  {journal} {\bibinfo  {journal} {Physical review letters}\ }\textbf {\bibinfo
  {volume} {109}},\ \bibinfo {pages} {096602} (\bibinfo {year}
  {2012}{\natexlab{b}})}\BibitemShut {NoStop}%
\bibitem [{\citenamefont {Manchon}\ \emph {et~al.}(2019)\citenamefont
  {Manchon}, \citenamefont {{\v{Z}}elezn{\`y}}, \citenamefont {Miron},
  \citenamefont {Jungwirth}, \citenamefont {Sinova}, \citenamefont {Thiaville},
  \citenamefont {Garello},\ and\ \citenamefont
  {Gambardella}}]{manchon2019current}%
  \BibitemOpen
  \bibfield  {author} {\bibinfo {author} {\bibfnamefont {A.}~\bibnamefont
  {Manchon}}, \bibinfo {author} {\bibfnamefont {J.}~\bibnamefont
  {{\v{Z}}elezn{\`y}}}, \bibinfo {author} {\bibfnamefont {I.~M.}\ \bibnamefont
  {Miron}}, \bibinfo {author} {\bibfnamefont {T.}~\bibnamefont {Jungwirth}},
  \bibinfo {author} {\bibfnamefont {J.}~\bibnamefont {Sinova}}, \bibinfo
  {author} {\bibfnamefont {A.}~\bibnamefont {Thiaville}}, \bibinfo {author}
  {\bibfnamefont {K.}~\bibnamefont {Garello}},\ and\ \bibinfo {author}
  {\bibfnamefont {P.}~\bibnamefont {Gambardella}},\ }\href@noop {} {\bibfield
  {journal} {\bibinfo  {journal} {Reviews of Modern Physics}\ }\textbf
  {\bibinfo {volume} {91}},\ \bibinfo {pages} {035004} (\bibinfo {year}
  {2019})}\BibitemShut {NoStop}%
\bibitem [{\citenamefont {Wang}\ \emph {et~al.}(2013)\citenamefont {Wang},
  \citenamefont {Alzate},\ and\ \citenamefont {Amiri}}]{wang2013low}%
  \BibitemOpen
  \bibfield  {author} {\bibinfo {author} {\bibfnamefont {K.}~\bibnamefont
  {Wang}}, \bibinfo {author} {\bibfnamefont {J.}~\bibnamefont {Alzate}},\ and\
  \bibinfo {author} {\bibfnamefont {P.~K.}\ \bibnamefont {Amiri}},\ }\href@noop
  {} {\bibfield  {journal} {\bibinfo  {journal} {Journal of Physics D: Applied
  Physics}\ }\textbf {\bibinfo {volume} {46}},\ \bibinfo {pages} {074003}
  (\bibinfo {year} {2013})}\BibitemShut {NoStop}%
\bibitem [{\citenamefont {Yu}\ \emph {et~al.}(2014)\citenamefont {Yu},
  \citenamefont {Upadhyaya}, \citenamefont {Fan}, \citenamefont {Alzate},
  \citenamefont {Jiang}, \citenamefont {Wong}, \citenamefont {Takei},
  \citenamefont {Bender}, \citenamefont {Chang}, \citenamefont {Jiang} \emph
  {et~al.}}]{yu2014switching}%
  \BibitemOpen
  \bibfield  {author} {\bibinfo {author} {\bibfnamefont {G.}~\bibnamefont
  {Yu}}, \bibinfo {author} {\bibfnamefont {P.}~\bibnamefont {Upadhyaya}},
  \bibinfo {author} {\bibfnamefont {Y.}~\bibnamefont {Fan}}, \bibinfo {author}
  {\bibfnamefont {J.~G.}\ \bibnamefont {Alzate}}, \bibinfo {author}
  {\bibfnamefont {W.}~\bibnamefont {Jiang}}, \bibinfo {author} {\bibfnamefont
  {K.~L.}\ \bibnamefont {Wong}}, \bibinfo {author} {\bibfnamefont
  {S.}~\bibnamefont {Takei}}, \bibinfo {author} {\bibfnamefont {S.~A.}\
  \bibnamefont {Bender}}, \bibinfo {author} {\bibfnamefont {L.-T.}\
  \bibnamefont {Chang}}, \bibinfo {author} {\bibfnamefont {Y.}~\bibnamefont
  {Jiang}}, \emph {et~al.},\ }\href@noop {} {\bibfield  {journal} {\bibinfo
  {journal} {Nature nanotechnology}\ }\textbf {\bibinfo {volume} {9}},\
  \bibinfo {pages} {548} (\bibinfo {year} {2014})}\BibitemShut {NoStop}%
\bibitem [{\citenamefont {MacNeill}\ \emph {et~al.}(2016)\citenamefont
  {MacNeill}, \citenamefont {Stiehl}, \citenamefont {Guimaraes}, \citenamefont
  {Buhrman}, \citenamefont {Park},\ and\ \citenamefont {Ralph}}]{MacNeill2016}%
  \BibitemOpen
  \bibfield  {author} {\bibinfo {author} {\bibfnamefont {D.}~\bibnamefont
  {MacNeill}}, \bibinfo {author} {\bibfnamefont {G.~M.}\ \bibnamefont
  {Stiehl}}, \bibinfo {author} {\bibfnamefont {M.~H.~D.}\ \bibnamefont
  {Guimaraes}}, \bibinfo {author} {\bibfnamefont {R.~A.}\ \bibnamefont
  {Buhrman}}, \bibinfo {author} {\bibfnamefont {J.}~\bibnamefont {Park}},\ and\
  \bibinfo {author} {\bibfnamefont {D.~C.}\ \bibnamefont {Ralph}},\ }\href
  {https://doi.org/10.1038/nphys3933} {\bibfield  {journal} {\bibinfo
  {journal} {Nature Physics}\ }\textbf {\bibinfo {volume} {13}},\ \bibinfo
  {pages} {300} (\bibinfo {year} {2016})}\BibitemShut {NoStop}%
\bibitem [{\citenamefont {MacNeill}\ \emph {et~al.}(2017)\citenamefont
  {MacNeill}, \citenamefont {Stiehl}, \citenamefont {Guimar\~aes},
  \citenamefont {Reynolds}, \citenamefont {Buhrman},\ and\ \citenamefont
  {Ralph}}]{MacNeill2017}%
  \BibitemOpen
  \bibfield  {author} {\bibinfo {author} {\bibfnamefont {D.}~\bibnamefont
  {MacNeill}}, \bibinfo {author} {\bibfnamefont {G.~M.}\ \bibnamefont
  {Stiehl}}, \bibinfo {author} {\bibfnamefont {M.~H.~D.}\ \bibnamefont
  {Guimar\~aes}}, \bibinfo {author} {\bibfnamefont {N.~D.}\ \bibnamefont
  {Reynolds}}, \bibinfo {author} {\bibfnamefont {R.~A.}\ \bibnamefont
  {Buhrman}},\ and\ \bibinfo {author} {\bibfnamefont {D.~C.}\ \bibnamefont
  {Ralph}},\ }\href {https://doi.org/10.1103/PhysRevB.96.054450} {\bibfield
  {journal} {\bibinfo  {journal} {Phys. Rev. B}\ }\textbf {\bibinfo {volume}
  {96}},\ \bibinfo {pages} {054450} (\bibinfo {year} {2017})}\BibitemShut
  {NoStop}%
\bibitem [{\citenamefont {Stiehl}\ \emph
  {et~al.}(2019{\natexlab{a}})\citenamefont {Stiehl}, \citenamefont {MacNeill},
  \citenamefont {Sivadas}, \citenamefont {El~Baggari}, \citenamefont
  {Guimaraes}, \citenamefont {Reynolds}, \citenamefont {Kourkoutis},
  \citenamefont {Fennie}, \citenamefont {Buhrman},\ and\ \citenamefont
  {Ralph}}]{stiehl2019current}%
  \BibitemOpen
  \bibfield  {author} {\bibinfo {author} {\bibfnamefont {G.~M.}\ \bibnamefont
  {Stiehl}}, \bibinfo {author} {\bibfnamefont {D.}~\bibnamefont {MacNeill}},
  \bibinfo {author} {\bibfnamefont {N.}~\bibnamefont {Sivadas}}, \bibinfo
  {author} {\bibfnamefont {I.}~\bibnamefont {El~Baggari}}, \bibinfo {author}
  {\bibfnamefont {M.~H.}\ \bibnamefont {Guimaraes}}, \bibinfo {author}
  {\bibfnamefont {N.~D.}\ \bibnamefont {Reynolds}}, \bibinfo {author}
  {\bibfnamefont {L.~F.}\ \bibnamefont {Kourkoutis}}, \bibinfo {author}
  {\bibfnamefont {C.~J.}\ \bibnamefont {Fennie}}, \bibinfo {author}
  {\bibfnamefont {R.~A.}\ \bibnamefont {Buhrman}},\ and\ \bibinfo {author}
  {\bibfnamefont {D.~C.}\ \bibnamefont {Ralph}},\ }\href@noop {} {\bibfield
  {journal} {\bibinfo  {journal} {ACS nano}\ }\textbf {\bibinfo {volume}
  {13}},\ \bibinfo {pages} {2599} (\bibinfo {year}
  {2019}{\natexlab{a}})}\BibitemShut {NoStop}%
\bibitem [{\citenamefont {Stiehl}\ \emph
  {et~al.}(2019{\natexlab{b}})\citenamefont {Stiehl}, \citenamefont {Li},
  \citenamefont {Gupta}, \citenamefont {Baggari}, \citenamefont {Jiang},
  \citenamefont {Xie}, \citenamefont {Kourkoutis}, \citenamefont {Mak},
  \citenamefont {Shan}, \citenamefont {Buhrman},\ and\ \citenamefont
  {Ralph}}]{Stiehl2019}%
  \BibitemOpen
  \bibfield  {author} {\bibinfo {author} {\bibfnamefont {G.~M.}\ \bibnamefont
  {Stiehl}}, \bibinfo {author} {\bibfnamefont {R.}~\bibnamefont {Li}}, \bibinfo
  {author} {\bibfnamefont {V.}~\bibnamefont {Gupta}}, \bibinfo {author}
  {\bibfnamefont {I.~E.}\ \bibnamefont {Baggari}}, \bibinfo {author}
  {\bibfnamefont {S.}~\bibnamefont {Jiang}}, \bibinfo {author} {\bibfnamefont
  {H.}~\bibnamefont {Xie}}, \bibinfo {author} {\bibfnamefont {L.~F.}\
  \bibnamefont {Kourkoutis}}, \bibinfo {author} {\bibfnamefont {K.~F.}\
  \bibnamefont {Mak}}, \bibinfo {author} {\bibfnamefont {J.}~\bibnamefont
  {Shan}}, \bibinfo {author} {\bibfnamefont {R.~A.}\ \bibnamefont {Buhrman}},\
  and\ \bibinfo {author} {\bibfnamefont {D.~C.}\ \bibnamefont {Ralph}},\ }\href
  {https://doi.org/10.1103/PhysRevB.100.184402} {\bibfield  {journal} {\bibinfo
   {journal} {Phys. Rev. B}\ }\textbf {\bibinfo {volume} {100}},\ \bibinfo
  {pages} {184402} (\bibinfo {year} {2019}{\natexlab{b}})}\BibitemShut
  {NoStop}%
\bibitem [{\citenamefont {Shi}\ \emph {et~al.}(2019)\citenamefont {Shi},
  \citenamefont {Liang}, \citenamefont {Zhu}, \citenamefont {Cai},
  \citenamefont {Pollard}, \citenamefont {Wang}, \citenamefont {Wang},
  \citenamefont {Wang}, \citenamefont {He}, \citenamefont {Yu} \emph
  {et~al.}}]{shi2019all}%
  \BibitemOpen
  \bibfield  {author} {\bibinfo {author} {\bibfnamefont {S.}~\bibnamefont
  {Shi}}, \bibinfo {author} {\bibfnamefont {S.}~\bibnamefont {Liang}}, \bibinfo
  {author} {\bibfnamefont {Z.}~\bibnamefont {Zhu}}, \bibinfo {author}
  {\bibfnamefont {K.}~\bibnamefont {Cai}}, \bibinfo {author} {\bibfnamefont
  {S.~D.}\ \bibnamefont {Pollard}}, \bibinfo {author} {\bibfnamefont
  {Y.}~\bibnamefont {Wang}}, \bibinfo {author} {\bibfnamefont {J.}~\bibnamefont
  {Wang}}, \bibinfo {author} {\bibfnamefont {Q.}~\bibnamefont {Wang}}, \bibinfo
  {author} {\bibfnamefont {P.}~\bibnamefont {He}}, \bibinfo {author}
  {\bibfnamefont {J.}~\bibnamefont {Yu}}, \emph {et~al.},\ }\href@noop {}
  {\bibfield  {journal} {\bibinfo  {journal} {Nature nanotechnology}\ }\textbf
  {\bibinfo {volume} {14}},\ \bibinfo {pages} {945} (\bibinfo {year}
  {2019})}\BibitemShut {NoStop}%
\bibitem [{\citenamefont {Xue}\ \emph {et~al.}(2020)\citenamefont {Xue},
  \citenamefont {Rohmann}, \citenamefont {Li}, \citenamefont {Amin},\ and\
  \citenamefont {Haney}}]{Xue2020SOT}%
  \BibitemOpen
  \bibfield  {author} {\bibinfo {author} {\bibfnamefont {F.}~\bibnamefont
  {Xue}}, \bibinfo {author} {\bibfnamefont {C.}~\bibnamefont {Rohmann}},
  \bibinfo {author} {\bibfnamefont {J.}~\bibnamefont {Li}}, \bibinfo {author}
  {\bibfnamefont {V.}~\bibnamefont {Amin}},\ and\ \bibinfo {author}
  {\bibfnamefont {P.}~\bibnamefont {Haney}},\ }\href
  {https://doi.org/10.1103/PhysRevB.102.014401} {\bibfield  {journal} {\bibinfo
   {journal} {Phys. Rev. B}\ }\textbf {\bibinfo {volume} {102}},\ \bibinfo
  {pages} {014401} (\bibinfo {year} {2020})}\BibitemShut {NoStop}%
\bibitem [{\citenamefont {Safeer}\ \emph {et~al.}(2019)\citenamefont {Safeer},
  \citenamefont {Ontoso}, \citenamefont {Ingla-Ayn{\'e}s}, \citenamefont
  {Herling}, \citenamefont {Pham}, \citenamefont {Kurzmann}, \citenamefont
  {Ensslin}, \citenamefont {Chuvilin}, \citenamefont {Robredo}, \citenamefont
  {Vergniory}, \citenamefont {de~Juan}, \citenamefont {Hueso}, \citenamefont
  {Calvo},\ and\ \citenamefont {Casanova}}]{Safeer2019}%
  \BibitemOpen
  \bibfield  {author} {\bibinfo {author} {\bibfnamefont {C.~K.}\ \bibnamefont
  {Safeer}}, \bibinfo {author} {\bibfnamefont {N.}~\bibnamefont {Ontoso}},
  \bibinfo {author} {\bibfnamefont {J.}~\bibnamefont {Ingla-Ayn{\'e}s}},
  \bibinfo {author} {\bibfnamefont {F.}~\bibnamefont {Herling}}, \bibinfo
  {author} {\bibfnamefont {V.~T.}\ \bibnamefont {Pham}}, \bibinfo {author}
  {\bibfnamefont {A.}~\bibnamefont {Kurzmann}}, \bibinfo {author}
  {\bibfnamefont {K.}~\bibnamefont {Ensslin}}, \bibinfo {author} {\bibfnamefont
  {A.}~\bibnamefont {Chuvilin}}, \bibinfo {author} {\bibfnamefont
  {I.}~\bibnamefont {Robredo}}, \bibinfo {author} {\bibfnamefont {M.~G.}\
  \bibnamefont {Vergniory}}, \bibinfo {author} {\bibfnamefont {F.}~\bibnamefont
  {de~Juan}}, \bibinfo {author} {\bibfnamefont {L.~E.}\ \bibnamefont {Hueso}},
  \bibinfo {author} {\bibfnamefont {M.~R.}\ \bibnamefont {Calvo}},\ and\
  \bibinfo {author} {\bibfnamefont {F.}~\bibnamefont {Casanova}},\ }\href
  {https://doi.org/10.1021/acs.nanolett.9b03485} {\bibfield  {journal}
  {\bibinfo  {journal} {Nano Letters}\ }\textbf {\bibinfo {volume} {19}},\
  \bibinfo {pages} {8758} (\bibinfo {year} {2019})}\BibitemShut {NoStop}%
\bibitem [{\citenamefont {Song}\ \emph {et~al.}(2020)\citenamefont {Song},
  \citenamefont {Hsu}, \citenamefont {Vignale}, \citenamefont {Zhao},
  \citenamefont {Liu}, \citenamefont {Deng}, \citenamefont {Fu}, \citenamefont
  {Liu}, \citenamefont {Zhang}, \citenamefont {Lin}, \citenamefont {Pereira},\
  and\ \citenamefont {Loh}}]{Song2020}%
  \BibitemOpen
  \bibfield  {author} {\bibinfo {author} {\bibfnamefont {P.}~\bibnamefont
  {Song}}, \bibinfo {author} {\bibfnamefont {C.-H.}\ \bibnamefont {Hsu}},
  \bibinfo {author} {\bibfnamefont {G.}~\bibnamefont {Vignale}}, \bibinfo
  {author} {\bibfnamefont {M.}~\bibnamefont {Zhao}}, \bibinfo {author}
  {\bibfnamefont {J.}~\bibnamefont {Liu}}, \bibinfo {author} {\bibfnamefont
  {Y.}~\bibnamefont {Deng}}, \bibinfo {author} {\bibfnamefont {W.}~\bibnamefont
  {Fu}}, \bibinfo {author} {\bibfnamefont {Y.}~\bibnamefont {Liu}}, \bibinfo
  {author} {\bibfnamefont {Y.}~\bibnamefont {Zhang}}, \bibinfo {author}
  {\bibfnamefont {H.}~\bibnamefont {Lin}}, \bibinfo {author} {\bibfnamefont
  {V.~M.}\ \bibnamefont {Pereira}},\ and\ \bibinfo {author} {\bibfnamefont
  {K.~P.}\ \bibnamefont {Loh}},\ }\href
  {https://doi.org/10.1038/s41563-019-0600-4} {\bibfield  {journal} {\bibinfo
  {journal} {Nature Materials}\ }\textbf {\bibinfo {volume} {19}},\ \bibinfo
  {pages} {292} (\bibinfo {year} {2020})}\BibitemShut {NoStop}%
\bibitem [{\citenamefont {Zhao}\ \emph {et~al.}()\citenamefont {Zhao},
  \citenamefont {Karpiak}, \citenamefont {Khokhriakov}, \citenamefont
  {Johansson}, \citenamefont {Hoque}, \citenamefont {Xu}, \citenamefont
  {Jiang}, \citenamefont {Mertig},\ and\ \citenamefont
  {Dash}}]{zhao2020unconventional}%
  \BibitemOpen
  \bibfield  {author} {\bibinfo {author} {\bibfnamefont {B.}~\bibnamefont
  {Zhao}}, \bibinfo {author} {\bibfnamefont {B.}~\bibnamefont {Karpiak}},
  \bibinfo {author} {\bibfnamefont {D.}~\bibnamefont {Khokhriakov}}, \bibinfo
  {author} {\bibfnamefont {A.}~\bibnamefont {Johansson}}, \bibinfo {author}
  {\bibfnamefont {A.~M.}\ \bibnamefont {Hoque}}, \bibinfo {author}
  {\bibfnamefont {X.}~\bibnamefont {Xu}}, \bibinfo {author} {\bibfnamefont
  {Y.}~\bibnamefont {Jiang}}, \bibinfo {author} {\bibfnamefont
  {I.}~\bibnamefont {Mertig}},\ and\ \bibinfo {author} {\bibfnamefont {S.~P.}\
  \bibnamefont {Dash}},\ }\href {https://doi.org/10.1002/adma.202000818}
  {\bibfield  {journal} {\bibinfo  {journal} {Advanced Materials}\ }\textbf
  {\bibinfo {volume} {n/a}},\ \bibinfo {pages} {2000818}},\ \Eprint
  {https://arxiv.org/abs/https://onlinelibrary.wiley.com/doi/pdf/10.1002/adma.202000818}
  {https://onlinelibrary.wiley.com/doi/pdf/10.1002/adma.202000818} \BibitemShut
  {NoStop}%
\bibitem [{\citenamefont {Seemann}\ \emph {et~al.}(2015)\citenamefont
  {Seemann}, \citenamefont {K\"odderitzsch}, \citenamefont {Wimmer},\ and\
  \citenamefont {Ebert}}]{Seemann2015}%
  \BibitemOpen
  \bibfield  {author} {\bibinfo {author} {\bibfnamefont {M.}~\bibnamefont
  {Seemann}}, \bibinfo {author} {\bibfnamefont {D.}~\bibnamefont
  {K\"odderitzsch}}, \bibinfo {author} {\bibfnamefont {S.}~\bibnamefont
  {Wimmer}},\ and\ \bibinfo {author} {\bibfnamefont {H.}~\bibnamefont
  {Ebert}},\ }\href {https://doi.org/10.1103/PhysRevB.92.155138} {\bibfield
  {journal} {\bibinfo  {journal} {Phys. Rev. B}\ }\textbf {\bibinfo {volume}
  {92}},\ \bibinfo {pages} {155138} (\bibinfo {year} {2015})}\BibitemShut
  {NoStop}%
\bibitem [{\citenamefont {Zhang}\ \emph {et~al.}(2014)\citenamefont {Zhang},
  \citenamefont {Liu}, \citenamefont {Luo}, \citenamefont {Freeman},\ and\
  \citenamefont {Zunger}}]{zhang2014hidden}%
  \BibitemOpen
  \bibfield  {author} {\bibinfo {author} {\bibfnamefont {X.}~\bibnamefont
  {Zhang}}, \bibinfo {author} {\bibfnamefont {Q.}~\bibnamefont {Liu}}, \bibinfo
  {author} {\bibfnamefont {J.-W.}\ \bibnamefont {Luo}}, \bibinfo {author}
  {\bibfnamefont {A.~J.}\ \bibnamefont {Freeman}},\ and\ \bibinfo {author}
  {\bibfnamefont {A.}~\bibnamefont {Zunger}},\ }\href@noop {} {\bibfield
  {journal} {\bibinfo  {journal} {Nature Physics}\ }\textbf {\bibinfo {volume}
  {10}},\ \bibinfo {pages} {387} (\bibinfo {year} {2014})}\BibitemShut
  {NoStop}%
\bibitem [{\citenamefont {Wadley}\ \emph {et~al.}(2016)\citenamefont {Wadley},
  \citenamefont {Howells}, \citenamefont {{\v{Z}}elezn{\`y}}, \citenamefont
  {Andrews}, \citenamefont {Hills}, \citenamefont {Campion}, \citenamefont
  {Nov{\'a}k}, \citenamefont {Olejn{\'\i}k}, \citenamefont {Maccherozzi},
  \citenamefont {Dhesi} \emph {et~al.}}]{wadley2016electrical}%
  \BibitemOpen
  \bibfield  {author} {\bibinfo {author} {\bibfnamefont {P.}~\bibnamefont
  {Wadley}}, \bibinfo {author} {\bibfnamefont {B.}~\bibnamefont {Howells}},
  \bibinfo {author} {\bibfnamefont {J.}~\bibnamefont {{\v{Z}}elezn{\`y}}},
  \bibinfo {author} {\bibfnamefont {C.}~\bibnamefont {Andrews}}, \bibinfo
  {author} {\bibfnamefont {V.}~\bibnamefont {Hills}}, \bibinfo {author}
  {\bibfnamefont {R.~P.}\ \bibnamefont {Campion}}, \bibinfo {author}
  {\bibfnamefont {V.}~\bibnamefont {Nov{\'a}k}}, \bibinfo {author}
  {\bibfnamefont {K.}~\bibnamefont {Olejn{\'\i}k}}, \bibinfo {author}
  {\bibfnamefont {F.}~\bibnamefont {Maccherozzi}}, \bibinfo {author}
  {\bibfnamefont {S.}~\bibnamefont {Dhesi}}, \emph {et~al.},\ }\href@noop {}
  {\bibfield  {journal} {\bibinfo  {journal} {Science}\ }\textbf {\bibinfo
  {volume} {351}},\ \bibinfo {pages} {587} (\bibinfo {year}
  {2016})}\BibitemShut {NoStop}%
\bibitem [{\citenamefont {Guo}\ \emph {et~al.}(2008)\citenamefont {Guo},
  \citenamefont {Murakami}, \citenamefont {Chen},\ and\ \citenamefont
  {Nagaosa}}]{Nagaosa_SHE}%
  \BibitemOpen
  \bibfield  {author} {\bibinfo {author} {\bibfnamefont {G.~Y.}\ \bibnamefont
  {Guo}}, \bibinfo {author} {\bibfnamefont {S.}~\bibnamefont {Murakami}},
  \bibinfo {author} {\bibfnamefont {T.-W.}\ \bibnamefont {Chen}},\ and\
  \bibinfo {author} {\bibfnamefont {N.}~\bibnamefont {Nagaosa}},\ }\href
  {https://doi.org/10.1103/PhysRevLett.100.096401} {\bibfield  {journal}
  {\bibinfo  {journal} {Phys. Rev. Lett.}\ }\textbf {\bibinfo {volume} {100}},\
  \bibinfo {pages} {096401} (\bibinfo {year} {2008})}\BibitemShut {NoStop}%
\bibitem [{\citenamefont {Tanaka}\ \emph {et~al.}(2008)\citenamefont {Tanaka},
  \citenamefont {Kontani}, \citenamefont {Naito}, \citenamefont {Naito},
  \citenamefont {Hirashima}, \citenamefont {Yamada},\ and\ \citenamefont
  {Inoue}}]{Tanaka_SHE}%
  \BibitemOpen
  \bibfield  {author} {\bibinfo {author} {\bibfnamefont {T.}~\bibnamefont
  {Tanaka}}, \bibinfo {author} {\bibfnamefont {H.}~\bibnamefont {Kontani}},
  \bibinfo {author} {\bibfnamefont {M.}~\bibnamefont {Naito}}, \bibinfo
  {author} {\bibfnamefont {T.}~\bibnamefont {Naito}}, \bibinfo {author}
  {\bibfnamefont {D.~S.}\ \bibnamefont {Hirashima}}, \bibinfo {author}
  {\bibfnamefont {K.}~\bibnamefont {Yamada}},\ and\ \bibinfo {author}
  {\bibfnamefont {J.}~\bibnamefont {Inoue}},\ }\href
  {https://doi.org/10.1103/PhysRevB.77.165117} {\bibfield  {journal} {\bibinfo
  {journal} {Phys. Rev. B}\ }\textbf {\bibinfo {volume} {77}},\ \bibinfo
  {pages} {165117} (\bibinfo {year} {2008})}\BibitemShut {NoStop}%
\bibitem [{\citenamefont {Todorov}(2002)}]{todorov2002tight}%
  \BibitemOpen
  \bibfield  {author} {\bibinfo {author} {\bibfnamefont {T.~N.}\ \bibnamefont
  {Todorov}},\ }\href@noop {} {\bibfield  {journal} {\bibinfo  {journal}
  {Journal of Physics: Condensed Matter}\ }\textbf {\bibinfo {volume} {14}},\
  \bibinfo {pages} {3049} (\bibinfo {year} {2002})}\BibitemShut {NoStop}%
\bibitem [{\citenamefont {Litvin}\ and\ \citenamefont
  {Kopsk{\'{y}}}(2011)}]{Litvin2011}%
  \BibitemOpen
  \bibfield  {author} {\bibinfo {author} {\bibfnamefont {D.~B.}\ \bibnamefont
  {Litvin}}\ and\ \bibinfo {author} {\bibfnamefont {V.}~\bibnamefont
  {Kopsk{\'{y}}}},\ }\href {https://doi.org/10.1107/S010876731101378X}
  {\bibfield  {journal} {\bibinfo  {journal} {Acta Crystallographica Section
  A}\ }\textbf {\bibinfo {volume} {67}},\ \bibinfo {pages} {415} (\bibinfo
  {year} {2011})}\BibitemShut {NoStop}%
\bibitem [{\citenamefont {Haney}\ and\ \citenamefont
  {Stiles}(2010)}]{haney2010current}%
  \BibitemOpen
  \bibfield  {author} {\bibinfo {author} {\bibfnamefont {P.~M.}\ \bibnamefont
  {Haney}}\ and\ \bibinfo {author} {\bibfnamefont {M.~D.}\ \bibnamefont
  {Stiles}},\ }\href@noop {} {\bibfield  {journal} {\bibinfo  {journal}
  {Physical review letters}\ }\textbf {\bibinfo {volume} {105}},\ \bibinfo
  {pages} {126602} (\bibinfo {year} {2010})}\BibitemShut {NoStop}%
\bibitem [{\citenamefont {Go}\ \emph {et~al.}(2020)\citenamefont {Go},
  \citenamefont {Freimuth}, \citenamefont {Hanke}, \citenamefont {Xue},
  \citenamefont {Gomonay}, \citenamefont {Lee}, \citenamefont {Blügel},
  \citenamefont {Lee},\ and\ \citenamefont {Mokrousov}}]{Go2020}%
  \BibitemOpen
  \bibfield  {author} {\bibinfo {author} {\bibfnamefont {D.}~\bibnamefont
  {Go}}, \bibinfo {author} {\bibfnamefont {F.}~\bibnamefont {Freimuth}},
  \bibinfo {author} {\bibfnamefont {J.-P.}\ \bibnamefont {Hanke}}, \bibinfo
  {author} {\bibfnamefont {F.}~\bibnamefont {Xue}}, \bibinfo {author}
  {\bibfnamefont {O.}~\bibnamefont {Gomonay}}, \bibinfo {author} {\bibfnamefont
  {K.-J.}\ \bibnamefont {Lee}}, \bibinfo {author} {\bibfnamefont
  {S.}~\bibnamefont {Blügel}}, \bibinfo {author} {\bibfnamefont {H.-W.}\
  \bibnamefont {Lee}},\ and\ \bibinfo {author} {\bibfnamefont {Y.}~\bibnamefont
  {Mokrousov}},\ }\href@noop {} {\bibinfo {title} {First-principles theory of
  current-induced spin-orbital coupled dynamics in magnetic heterostructures}}
  (\bibinfo {year} {2020}),\ \Eprint {https://arxiv.org/abs/2004.05945}
  {arXiv:2004.05945 [cond-mat.mes-hall]} \BibitemShut {NoStop}%
\bibitem [{\citenamefont {Soluyanov}\ \emph {et~al.}(2015)\citenamefont
  {Soluyanov}, \citenamefont {Gresch}, \citenamefont {Wang}, \citenamefont
  {Wu}, \citenamefont {Troyer}, \citenamefont {Dai},\ and\ \citenamefont
  {Bernevig}}]{Soluyanov2015}%
  \BibitemOpen
  \bibfield  {author} {\bibinfo {author} {\bibfnamefont {A.~A.}\ \bibnamefont
  {Soluyanov}}, \bibinfo {author} {\bibfnamefont {D.}~\bibnamefont {Gresch}},
  \bibinfo {author} {\bibfnamefont {Z.}~\bibnamefont {Wang}}, \bibinfo {author}
  {\bibfnamefont {Q.}~\bibnamefont {Wu}}, \bibinfo {author} {\bibfnamefont
  {M.}~\bibnamefont {Troyer}}, \bibinfo {author} {\bibfnamefont
  {X.}~\bibnamefont {Dai}},\ and\ \bibinfo {author} {\bibfnamefont {B.~A.}\
  \bibnamefont {Bernevig}},\ }\href {https://doi.org/10.1038/nature15768}
  {\bibfield  {journal} {\bibinfo  {journal} {Nature}\ }\textbf {\bibinfo
  {volume} {527}},\ \bibinfo {pages} {495} (\bibinfo {year}
  {2015})}\BibitemShut {NoStop}%
\bibitem [{\citenamefont {Zhou}\ \emph {et~al.}(2019)\citenamefont {Zhou},
  \citenamefont {Qiao}, \citenamefont {Bournel},\ and\ \citenamefont
  {Zhao}}]{Zhou2019}%
  \BibitemOpen
  \bibfield  {author} {\bibinfo {author} {\bibfnamefont {J.}~\bibnamefont
  {Zhou}}, \bibinfo {author} {\bibfnamefont {J.}~\bibnamefont {Qiao}}, \bibinfo
  {author} {\bibfnamefont {A.}~\bibnamefont {Bournel}},\ and\ \bibinfo {author}
  {\bibfnamefont {W.}~\bibnamefont {Zhao}},\ }\href
  {https://doi.org/10.1103/PhysRevB.99.060408} {\bibfield  {journal} {\bibinfo
  {journal} {Phys. Rev. B}\ }\textbf {\bibinfo {volume} {99}},\ \bibinfo
  {pages} {060408} (\bibinfo {year} {2019})}\BibitemShut {NoStop}%
\bibitem [{\citenamefont {Zhao}\ \emph {et~al.}(2020)\citenamefont {Zhao},
  \citenamefont {Khokhriakov}, \citenamefont {Zhang}, \citenamefont {Fu},
  \citenamefont {Karpiak}, \citenamefont {Hoque}, \citenamefont {Xu},
  \citenamefont {Jiang}, \citenamefont {Yan},\ and\ \citenamefont
  {Dash}}]{Dash2020Conventional}%
  \BibitemOpen
  \bibfield  {author} {\bibinfo {author} {\bibfnamefont {B.}~\bibnamefont
  {Zhao}}, \bibinfo {author} {\bibfnamefont {D.}~\bibnamefont {Khokhriakov}},
  \bibinfo {author} {\bibfnamefont {Y.}~\bibnamefont {Zhang}}, \bibinfo
  {author} {\bibfnamefont {H.}~\bibnamefont {Fu}}, \bibinfo {author}
  {\bibfnamefont {B.}~\bibnamefont {Karpiak}}, \bibinfo {author} {\bibfnamefont
  {A.~M.}\ \bibnamefont {Hoque}}, \bibinfo {author} {\bibfnamefont
  {X.}~\bibnamefont {Xu}}, \bibinfo {author} {\bibfnamefont {Y.}~\bibnamefont
  {Jiang}}, \bibinfo {author} {\bibfnamefont {B.}~\bibnamefont {Yan}},\ and\
  \bibinfo {author} {\bibfnamefont {S.~P.}\ \bibnamefont {Dash}},\ }\href
  {https://doi.org/10.1103/PhysRevResearch.2.013286} {\bibfield  {journal}
  {\bibinfo  {journal} {Phys. Rev. Research}\ }\textbf {\bibinfo {volume}
  {2}},\ \bibinfo {pages} {013286} (\bibinfo {year} {2020})}\BibitemShut
  {NoStop}%
\bibitem [{\citenamefont {Mariano}\ and\ \citenamefont
  {Chopra}(1967)}]{PbTe1967}%
  \BibitemOpen
  \bibfield  {author} {\bibinfo {author} {\bibfnamefont {A.~N.}\ \bibnamefont
  {Mariano}}\ and\ \bibinfo {author} {\bibfnamefont {K.~L.}\ \bibnamefont
  {Chopra}},\ }\href {https://doi.org/10.1063/1.1754812} {\bibfield  {journal}
  {\bibinfo  {journal} {Applied Physics Letters}\ }\textbf {\bibinfo {volume}
  {10}},\ \bibinfo {pages} {282} (\bibinfo {year} {1967})},\ \Eprint
  {https://arxiv.org/abs/https://doi.org/10.1063/1.1754812}
  {https://doi.org/10.1063/1.1754812} \BibitemShut {NoStop}%
\bibitem [{\citenamefont {Rauch}\ \emph {et~al.}(2020)\citenamefont {Rauch},
  \citenamefont {T\"opler},\ and\ \citenamefont {Mertig}}]{Tomas2020}%
  \BibitemOpen
  \bibfield  {author} {\bibinfo {author} {\bibfnamefont {T.}~\bibnamefont
  {Rauch}}, \bibinfo {author} {\bibfnamefont {F.}~\bibnamefont {T\"opler}},\
  and\ \bibinfo {author} {\bibfnamefont {I.}~\bibnamefont {Mertig}},\ }\href
  {https://doi.org/10.1103/PhysRevB.101.064206} {\bibfield  {journal} {\bibinfo
   {journal} {Phys. Rev. B}\ }\textbf {\bibinfo {volume} {101}},\ \bibinfo
  {pages} {064206} (\bibinfo {year} {2020})}\BibitemShut {NoStop}%
\bibitem [{\citenamefont {Giannozzi}\ \emph {et~al.}(2017)\citenamefont
  {Giannozzi}, \citenamefont {Andreussi}, \citenamefont {Brumme}, \citenamefont
  {Bunau}, \citenamefont {Nardelli}, \citenamefont {Calandra}, \citenamefont
  {Car}, \citenamefont {Cavazzoni}, \citenamefont {Ceresoli}, \citenamefont
  {Cococcioni}, \citenamefont {Colonna}, \citenamefont {Carnimeo},
  \citenamefont {Corso}, \citenamefont {de~Gironcoli}, \citenamefont {Delugas},
  \citenamefont {DiStasio}, \citenamefont {Ferretti}, \citenamefont {Floris},
  \citenamefont {Fratesi}, \citenamefont {Fugallo}, \citenamefont {Gebauer},
  \citenamefont {Gerstmann}, \citenamefont {Giustino}, \citenamefont {Gorni},
  \citenamefont {Jia}, \citenamefont {Kawamura}, \citenamefont {Ko},
  \citenamefont {Kokalj}, \citenamefont {Kü{\c{c}}ükbenli}, \citenamefont
  {Lazzeri}, \citenamefont {Marsili}, \citenamefont {Marzari}, \citenamefont
  {Mauri}, \citenamefont {Nguyen}, \citenamefont {Nguyen}, \citenamefont {de-la
  Roza}, \citenamefont {Paulatto}, \citenamefont {Ponc{\'{e}}}, \citenamefont
  {Rocca}, \citenamefont {Sabatini}, \citenamefont {Santra}, \citenamefont
  {Schlipf}, \citenamefont {Seitsonen}, \citenamefont {Smogunov}, \citenamefont
  {Timrov}, \citenamefont {Thonhauser}, \citenamefont {Umari}, \citenamefont
  {Vast}, \citenamefont {Wu},\ and\ \citenamefont {Baroni}}]{QE}%
  \BibitemOpen
  \bibfield  {author} {\bibinfo {author} {\bibfnamefont {P.}~\bibnamefont
  {Giannozzi}}, \bibinfo {author} {\bibfnamefont {O.}~\bibnamefont
  {Andreussi}}, \bibinfo {author} {\bibfnamefont {T.}~\bibnamefont {Brumme}},
  \bibinfo {author} {\bibfnamefont {O.}~\bibnamefont {Bunau}}, \bibinfo
  {author} {\bibfnamefont {M.~B.}\ \bibnamefont {Nardelli}}, \bibinfo {author}
  {\bibfnamefont {M.}~\bibnamefont {Calandra}}, \bibinfo {author}
  {\bibfnamefont {R.}~\bibnamefont {Car}}, \bibinfo {author} {\bibfnamefont
  {C.}~\bibnamefont {Cavazzoni}}, \bibinfo {author} {\bibfnamefont
  {D.}~\bibnamefont {Ceresoli}}, \bibinfo {author} {\bibfnamefont
  {M.}~\bibnamefont {Cococcioni}}, \bibinfo {author} {\bibfnamefont
  {N.}~\bibnamefont {Colonna}}, \bibinfo {author} {\bibfnamefont
  {I.}~\bibnamefont {Carnimeo}}, \bibinfo {author} {\bibfnamefont {A.~D.}\
  \bibnamefont {Corso}}, \bibinfo {author} {\bibfnamefont {S.}~\bibnamefont
  {de~Gironcoli}}, \bibinfo {author} {\bibfnamefont {P.}~\bibnamefont
  {Delugas}}, \bibinfo {author} {\bibfnamefont {R.~A.}\ \bibnamefont
  {DiStasio}}, \bibinfo {author} {\bibfnamefont {A.}~\bibnamefont {Ferretti}},
  \bibinfo {author} {\bibfnamefont {A.}~\bibnamefont {Floris}}, \bibinfo
  {author} {\bibfnamefont {G.}~\bibnamefont {Fratesi}}, \bibinfo {author}
  {\bibfnamefont {G.}~\bibnamefont {Fugallo}}, \bibinfo {author} {\bibfnamefont
  {R.}~\bibnamefont {Gebauer}}, \bibinfo {author} {\bibfnamefont
  {U.}~\bibnamefont {Gerstmann}}, \bibinfo {author} {\bibfnamefont
  {F.}~\bibnamefont {Giustino}}, \bibinfo {author} {\bibfnamefont
  {T.}~\bibnamefont {Gorni}}, \bibinfo {author} {\bibfnamefont
  {J.}~\bibnamefont {Jia}}, \bibinfo {author} {\bibfnamefont {M.}~\bibnamefont
  {Kawamura}}, \bibinfo {author} {\bibfnamefont {H.-Y.}\ \bibnamefont {Ko}},
  \bibinfo {author} {\bibfnamefont {A.}~\bibnamefont {Kokalj}}, \bibinfo
  {author} {\bibfnamefont {E.}~\bibnamefont {Kü{\c{c}}ükbenli}}, \bibinfo
  {author} {\bibfnamefont {M.}~\bibnamefont {Lazzeri}}, \bibinfo {author}
  {\bibfnamefont {M.}~\bibnamefont {Marsili}}, \bibinfo {author} {\bibfnamefont
  {N.}~\bibnamefont {Marzari}}, \bibinfo {author} {\bibfnamefont
  {F.}~\bibnamefont {Mauri}}, \bibinfo {author} {\bibfnamefont {N.~L.}\
  \bibnamefont {Nguyen}}, \bibinfo {author} {\bibfnamefont {H.-V.}\
  \bibnamefont {Nguyen}}, \bibinfo {author} {\bibfnamefont {A.~O.}\
  \bibnamefont {de-la Roza}}, \bibinfo {author} {\bibfnamefont
  {L.}~\bibnamefont {Paulatto}}, \bibinfo {author} {\bibfnamefont
  {S.}~\bibnamefont {Ponc{\'{e}}}}, \bibinfo {author} {\bibfnamefont
  {D.}~\bibnamefont {Rocca}}, \bibinfo {author} {\bibfnamefont
  {R.}~\bibnamefont {Sabatini}}, \bibinfo {author} {\bibfnamefont
  {B.}~\bibnamefont {Santra}}, \bibinfo {author} {\bibfnamefont
  {M.}~\bibnamefont {Schlipf}}, \bibinfo {author} {\bibfnamefont {A.~P.}\
  \bibnamefont {Seitsonen}}, \bibinfo {author} {\bibfnamefont {A.}~\bibnamefont
  {Smogunov}}, \bibinfo {author} {\bibfnamefont {I.}~\bibnamefont {Timrov}},
  \bibinfo {author} {\bibfnamefont {T.}~\bibnamefont {Thonhauser}}, \bibinfo
  {author} {\bibfnamefont {P.}~\bibnamefont {Umari}}, \bibinfo {author}
  {\bibfnamefont {N.}~\bibnamefont {Vast}}, \bibinfo {author} {\bibfnamefont
  {X.}~\bibnamefont {Wu}},\ and\ \bibinfo {author} {\bibfnamefont
  {S.}~\bibnamefont {Baroni}},\ }\href
  {https://doi.org/10.1088/1361-648x/aa8f79} {\bibfield  {journal} {\bibinfo
  {journal} {Journal of Physics: Condensed Matter}\ }\textbf {\bibinfo {volume}
  {29}},\ \bibinfo {pages} {465901} (\bibinfo {year} {2017})}\BibitemShut
  {NoStop}%
\bibitem [{\citenamefont {Mostofi}\ \emph {et~al.}(2014)\citenamefont
  {Mostofi}, \citenamefont {Yates}, \citenamefont {Pizzi}, \citenamefont {Lee},
  \citenamefont {Souza}, \citenamefont {Vanderbilt},\ and\ \citenamefont
  {Marzari}}]{Wannier90}%
  \BibitemOpen
  \bibfield  {author} {\bibinfo {author} {\bibfnamefont {A.~A.}\ \bibnamefont
  {Mostofi}}, \bibinfo {author} {\bibfnamefont {J.~R.}\ \bibnamefont {Yates}},
  \bibinfo {author} {\bibfnamefont {G.}~\bibnamefont {Pizzi}}, \bibinfo
  {author} {\bibfnamefont {Y.-S.}\ \bibnamefont {Lee}}, \bibinfo {author}
  {\bibfnamefont {I.}~\bibnamefont {Souza}}, \bibinfo {author} {\bibfnamefont
  {D.}~\bibnamefont {Vanderbilt}},\ and\ \bibinfo {author} {\bibfnamefont
  {N.}~\bibnamefont {Marzari}},\ }\href
  {https://doi.org/https://doi.org/10.1016/j.cpc.2014.05.003} {\bibfield
  {journal} {\bibinfo  {journal} {Computer Physics Communications}\ }\textbf
  {\bibinfo {volume} {185}},\ \bibinfo {pages} {2309 } (\bibinfo {year}
  {2014})}\BibitemShut {NoStop}%
\bibitem [{\citenamefont {Corso]}(2014)}]{DALCORSO2014337}%
  \BibitemOpen
  \bibfield  {author} {\bibinfo {author} {\bibfnamefont {A.~D.}\ \bibnamefont
  {Corso]}},\ }\href
  {https://doi.org/https://doi.org/10.1016/j.commatsci.2014.07.043} {\bibfield
  {journal} {\bibinfo  {journal} {Computational Materials Science}\ }\textbf
  {\bibinfo {volume} {95}},\ \bibinfo {pages} {337 } (\bibinfo {year}
  {2014})}\BibitemShut {NoStop}%
\bibitem [{\citenamefont {Kresse}\ and\ \citenamefont {Joubert}(1999)}]{PAW}%
  \BibitemOpen
  \bibfield  {author} {\bibinfo {author} {\bibfnamefont {G.}~\bibnamefont
  {Kresse}}\ and\ \bibinfo {author} {\bibfnamefont {D.}~\bibnamefont
  {Joubert}},\ }\href {https://doi.org/10.1103/PhysRevB.59.1758} {\bibfield
  {journal} {\bibinfo  {journal} {Phys. Rev. B}\ }\textbf {\bibinfo {volume}
  {59}},\ \bibinfo {pages} {1758} (\bibinfo {year} {1999})}\BibitemShut
  {NoStop}%
\bibitem [{\citenamefont {Perdew}\ \emph {et~al.}(1996)\citenamefont {Perdew},
  \citenamefont {Burke},\ and\ \citenamefont {Ernzerhof}}]{PBEGGA}%
  \BibitemOpen
  \bibfield  {author} {\bibinfo {author} {\bibfnamefont {J.~P.}\ \bibnamefont
  {Perdew}}, \bibinfo {author} {\bibfnamefont {K.}~\bibnamefont {Burke}},\ and\
  \bibinfo {author} {\bibfnamefont {M.}~\bibnamefont {Ernzerhof}},\ }\href
  {https://doi.org/10.1103/PhysRevLett.77.3865} {\bibfield  {journal} {\bibinfo
   {journal} {Phys. Rev. Lett.}\ }\textbf {\bibinfo {volume} {77}},\ \bibinfo
  {pages} {3865} (\bibinfo {year} {1996})}\BibitemShut {NoStop}%
\bibitem [{\citenamefont {Monkhorst}\ and\ \citenamefont
  {Pack}(1976)}]{MPmesh}%
  \BibitemOpen
  \bibfield  {author} {\bibinfo {author} {\bibfnamefont {H.~J.}\ \bibnamefont
  {Monkhorst}}\ and\ \bibinfo {author} {\bibfnamefont {J.~D.}\ \bibnamefont
  {Pack}},\ }\href {https://doi.org/10.1103/PhysRevB.13.5188} {\bibfield
  {journal} {\bibinfo  {journal} {Phys. Rev. B}\ }\textbf {\bibinfo {volume}
  {13}},\ \bibinfo {pages} {5188} (\bibinfo {year} {1976})}\BibitemShut
  {NoStop}%
\bibitem [{\citenamefont {Gresch}\ \emph {et~al.}(2018)\citenamefont {Gresch},
  \citenamefont {Wu}, \citenamefont {Winkler}, \citenamefont {H\"auselmann},
  \citenamefont {Troyer},\ and\ \citenamefont {Soluyanov}}]{TBmodels}%
  \BibitemOpen
  \bibfield  {author} {\bibinfo {author} {\bibfnamefont {D.}~\bibnamefont
  {Gresch}}, \bibinfo {author} {\bibfnamefont {Q.}~\bibnamefont {Wu}}, \bibinfo
  {author} {\bibfnamefont {G.~W.}\ \bibnamefont {Winkler}}, \bibinfo {author}
  {\bibfnamefont {R.}~\bibnamefont {H\"auselmann}}, \bibinfo {author}
  {\bibfnamefont {M.}~\bibnamefont {Troyer}},\ and\ \bibinfo {author}
  {\bibfnamefont {A.~A.}\ \bibnamefont {Soluyanov}},\ }\href
  {https://doi.org/10.1103/PhysRevMaterials.2.103805} {\bibfield  {journal}
  {\bibinfo  {journal} {Phys. Rev. Materials}\ }\textbf {\bibinfo {volume}
  {2}},\ \bibinfo {pages} {103805} (\bibinfo {year} {2018})}\BibitemShut
  {NoStop}%
\bibitem [{\citenamefont {Ryoo}\ \emph {et~al.}(2019)\citenamefont {Ryoo},
  \citenamefont {Park},\ and\ \citenamefont {Souza}}]{Souza2019}%
  \BibitemOpen
  \bibfield  {author} {\bibinfo {author} {\bibfnamefont {J.~H.}\ \bibnamefont
  {Ryoo}}, \bibinfo {author} {\bibfnamefont {C.-H.}\ \bibnamefont {Park}},\
  and\ \bibinfo {author} {\bibfnamefont {I.}~\bibnamefont {Souza}},\ }\href
  {https://doi.org/10.1103/PhysRevB.99.235113} {\bibfield  {journal} {\bibinfo
  {journal} {Phys. Rev. B}\ }\textbf {\bibinfo {volume} {99}},\ \bibinfo
  {pages} {235113} (\bibinfo {year} {2019})}\BibitemShut {NoStop}%
\end{thebibliography}%
\bibliographystyle{apsrev4-2}
\end{document}